\documentclass[lettersize,journal]{IEEEtran}

\usepackage{amsmath,amsfonts}
\usepackage{algorithm}
\usepackage{algpseudocode}
\usepackage{array}
\usepackage[caption=false,font=footnotesize]{subfig}
\usepackage{textcomp}
\usepackage{stfloats}
\usepackage{url}
\usepackage{verbatim}
\usepackage{graphicx}
\usepackage{cite}
\usepackage{bm} 
\usepackage{gensymb}
\usepackage{hyperref}
\usepackage{inputenc}
\usepackage{amsmath}
\usepackage{enumerate}
\usepackage{textgreek}
\usepackage{booktabs}
\usepackage{nicefrac}
\usepackage[T1]{fontenc}
\DeclareFontFamily{T1}{calligra}{}
\DeclareFontShape{T1}{calligra}{m}{n}{<->s*[1.44]callig15}{}
\DeclareMathAlphabet\mathcalligra   {T1}{calligra} {m} {n}
\DeclareMathAlphabet\mathzapf       {T1}{pzc} {mb} {it}
\DeclareMathAlphabet\mathchorus     {T1}{qzc} {m} {n}
\DeclareMathAlphabet\mathrsfso      {U}{rsfso}{m}{n}
\DeclareMathAlphabet\mathfrc{T1}{frc}{m}{it}
\usepackage{capt-of}  
\usepackage{cuted}
\usepackage{placeins}

\usepackage{xcolor}

\newcommand{\vertspacebeforesections}{-1mm}
\newcommand{\vertspacebeforesubsections}{-2mm}

\usepackage{cleveref}
\Crefname{figure}{Fig.}{Figs.} 

\usepackage{siunitx}

\usepackage{balance} 

\begin{document}
\title{Distributed Machine Learning Approach for Low-Latency Localization in Cell-Free Massive MIMO Systems}
\author{Manish Kumar, Tzu-Hsuan Chou, Byunghyun Lee, Nicolò Michelusi,\\David J. Love, Yaguang Zhang, and James V. Krogmeier
\thanks{
A preliminary version of this work has been submitted to the Asilomar Conference on Signals, Systems, and Computers, 2025~\cite{my_wcl_letter}.

M. Kumar, B. Lee, D. J. Love and J. V. Krogmeier are with the Elmore Family School of Electrical and Computer Engineering, Purdue University, West Lafayette, IN, USA; emails: \{mkrishne, lee4093, djlove, jvk\}@purdue.edu.}
\thanks{Y. Zhang is with the Department of Agricultural and Biological Engineering and the Department of Agricultural Sciences Education and Communication, Purdue University, West Lafayette, IN, USA; email: ygzhang@purdue.edu.}
\thanks{T.-H. Chou is with Qualcomm, Inc., San Diego, CA 92121 USA; email: tzuhchou@qti.qualcomm.com.}
\thanks{N. Michelusi is with the School of Electrical, Computer and Energy Engineering, Arizona State University, AZ, USA; email: nicolo.michelusi@asu.edu. His research is funded by NSF CNS-2129015.}
\thanks{This work is supported in part by the National Science Foundation under grants EEC-1941529, CNS-2235134, CNS-2212565 and CNS-2225578.}
\vspace{-8mm}
}
\maketitle
\begin{abstract}
Low-latency localization is critical in cellular networks to support real-time applications requiring precise positioning. In this paper, we propose a distributed machine learning (ML) framework for fingerprint-based localization tailored to cell-free massive multiple-input multiple-output (MIMO) systems, an emerging architecture for 6G networks. The proposed framework enables each access point (AP) to independently train a Gaussian process regression model using local angle-of-arrival and received signal strength fingerprints. These models provide probabilistic position estimates for the user equipment (UE), which are then fused by the UE with minimal computational overhead to derive a final location estimate. This decentralized approach eliminates the need for fronthaul communication between the APs and the central processing unit (CPU), thereby reducing latency. Additionally, distributing computational tasks across the APs alleviates the processing burden on the CPU compared to traditional centralized localization schemes. Simulation results demonstrate that the proposed distributed framework achieves localization accuracy comparable to centralized methods, despite lacking the benefits of centralized data aggregation. Moreover, it effectively reduces uncertainty of the location estimates, as evidenced by the 95\% covariance ellipse. The results highlight the potential of distributed ML for enabling low-latency, high-accuracy localization in future 6G networks. 
\end{abstract}


\vspace{-5mm}

\section{Introduction}
\label{introduction}
The next-generation 6G mobile communication is expected to revolutionize wireless communication systems, with integrated sensing and communication (ISAC) playing a key role in enabling advanced connectivity. ISAC integrates communication and sensing capabilities into a single framework, unlocking new opportunities for enhanced performance in areas such as positioning and localization~\cite{Love_6G_paper}. Accurate positioning and low-latency localization are essential for applications ranging from autonomous navigation to emergency response. The cell-free massive multiple-input multiple-output (MIMO) system, a key candidate for 6G~\cite{cell_free_as_6G}, utilizes multiple access points (APs) located closer to users, making it particularly well-suited for these location and latency critical applications.

The cooperation among the APs in cell-free massive MIMO systems can be coordinated through either centralized or distributed processing~\cite{cell_free_mimo_book}. In centralized operation, the APs mainly function as relays that forward baseband signals to a central processing unit (CPU). The CPU is responsible for channel estimation, data detection, and resource allocation. While this enables globally optimized decisions, it introduces significant latency due to fronthaul communication and high computational complexity at the CPU, especially for localization tasks~\cite{cell_free_mimo_book,fronthaul_quant,my_wcl_letter}. Additionally, signal quantization for fronthaul transmission introduces quantization distortion~\cite{fronthaul_quant}.

In contrast, distributed operation enables individual APs to perform most of the processing independently, thereby reducing fronthaul load and improving scalability. Building on this principle, we propose a localization framework \footnote{Although the terms are often used interchangeably, positioning typically refers to estimating exact coordinates, while localization identifies a region, relative to the environment, where the object is likely located. Since Gaussian process regression provides both position estimates and associated uncertainty regions (e.g. 95\% error ellipse), we refer to our approach as `localization'.} for cell-free massive MIMO systems that utilizes its distributed operation. In this framework, the APs compute the received signal strength (RSS) and angle-of-arrival (AOA) estimates from predefined points in the system,
referred to as reference points (RPs), and store this information in the AP's local database. Each AP then uses Gaussian process regression (GPR) as the underlying machine learning (ML) method to model location coordinates and independently generate position estimates, which are subsequently transmitted to the user equipment (UE) for final fusion. The key contributions of this work are summarized as follows:
\begin{enumerate}
    \item We propose an ML framework for UE localization in cell-free massive MIMO systems, which eliminates the need for fronthaul data exchange between APs and the CPU.
    \item We introduce a fully distributed learning approach where each AP independently trains a GPR model using local RSS and AOA measurements to estimate UE positions.
    \item We design a UE-assisted localization strategy, where the UE combines position estimates from multiple APs. Different methods for combining individual AP estimates are presented as the operational variants of the framework.
    \item We evaluate the proposed method against centralized localization techniques using the mean localization error, 95\% confidence ellipse and computational complexity, demonstrating comparable localization accuracy.
    \item We develop a simulation framework and conduct extensive evaluations under diverse system conditions, including varying antenna counts per AP, shadowing levels, and different AP and RP densities.
    \item We show that by achieving the Cramer-Rao bound (CRB) for AOA estimation, the localization accuracy can be further improved, motivating the development of more precise AOA estimation techniques.
\end{enumerate}

The remainder of this paper is structured as follows: \Cref{related_work} presents a literature review of positioning and localization methods for MIMO systems and positions our approach within the broader context of ML-based techniques. \Cref{system_model} describes the system model, while \Cref{fingerprint_extraction} details the strategy for RSS and AOA measurement. \Cref{distributed_loc_framework} introduces the distributed localization framework and its operational variants. In \Cref{complexity_and_performance}, the computational complexity of the proposed algorithms are analyzed and key performance metrics for evaluation are defined. \Cref{sim_results} presents simulation results and \Cref{conclusion} concludes the paper.

\textbf{Notation}: 
Matrices and vectors are represented by boldface uppercase and lowercase letters, respectively. The operations $(\cdot)^{\top}$, $(\cdot)^H$, and $(\cdot)^{-1}$ represent the transpose, conjugate transpose, and inverse, respectively. The notation ${col}_i(\cdot)$ refers to the $i$-th column of a matrix or the $i$-th element of a row vector, while ${row}_i(\cdot)$ refers to the $i$-th row of a matrix. The Euclidean and Frobenius norms are represented by $\|\cdot\|$ and $\|\cdot\|_F$ respectively. $\text{Tr}(\cdot)$ denotes the trace and $\text{diag}(\mathbf{v})$ is a diagonal matrix where the elements of vector $\mathbf{v}$ are placed along the principal diagonal. $[\mathbf{A}]_{mn}$ denotes the $(m,n)^{th}$ element of matrix $\mathbf{A}$. The expectation and gradient operators are denoted as $\mathbb{E}[\cdot]$ and $\nabla$ respectively. The notation $\mathbf{0}_{M\times N}$ represents an $M \times N$ zero matrix, and $\mathbf{I}_N$ is the $N \times N$ identity matrix. $\frac{\partial y}{\partial x}$ is the partial derivative of $y$ with respect to $x$. The Hadamard product is represented by $\odot$, and the Kronecker delta function $\delta_{pq}$ equals $1$ if $p = q$, and $0$ otherwise.

\vspace{\vertspacebeforesections}
\section{Related Work}
\label{related_work}
\subsection{Fingerprint-Based Positioning in MIMO Systems}
Fingerprint-based positioning leverages ML techniques such as supervised, unsupervised, and reinforcement learning to infer user positions from signal features~\cite{my_wcl_letter,joint_aoa_rss,cooperative_positioning1}. They benefit from rich positional data across multiple base stations and are robust to non-line-of-sight (NLOS) biases caused by multipath propagation. Unsupervised methods, like k-means clustering, affinity propagation and self-organizing maps, group similar fingerprints and adapt to environmental changes by continuously updating clusters~\cite{joint_aoa_rss,fog_massive_mimo_ml,cooperative_positioning1,sec_trans_positioning,gpr_plus_clustering}. Recently, reinforcement learning-based positioning has gained attention due to its adaptability to dynamic signal environments~\cite{rl_pos2}.

Supervised positioning algorithms use labeled data which pair signal measurements with known RP locations during training. This enables higher accuracy compared to unsupervised ML, making supervised methods ideal for fingerprint-based positioning. Linear-regression (LR), k-nearest neighbors (KNN)~\cite{fog_massive_mimo_ml,3D_WKNN_NN,knn_pos3}, neural networks~\cite{fog_massive_mimo_ml,distributed_music_positioning,3D_WKNN_NN}, and GPRs~\cite{fingerprint_savic,ml_rss_surya,analytical_surya,my_wcl_letter}, have been explored in massive MIMO research. We adopt GPR for its superior localization accuracy and its ability to provide probabilistic estimates~\cite{my_wcl_letter,ml_rss_surya}.

\vspace{\vertspacebeforesubsections}
\subsection{RSS- and AOA-Based Fingerprinting Techniques}
RSS is widely used in fingerprint-based positioning due to its simplicity and direct relation to distance~\cite{fingerprint_savic,ml_rss_surya,fog_massive_mimo_ml,analytical_surya}. However, in distributed positioning, RSS alone is insufficient for independent position estimation at each AP. To address this, it is necessary to incorporate a secondary fingerprint, such as AOA, time-of-arrival or time-difference-of-arrival, that can provide the AP with the required supplementary information. 

AOA fingerprints, in particular, do not require tight clock synchronization across APs and benefit from antenna array gains, making them suitable for large-scale cell-free massive MIMO systems. Existing works employ angle domain channel power matrices \cite{cooperative_positioning1,sec_trans_positioning,joint_aoa_rss} and multiple signal classification (MUSIC) based AOA estimates \cite{ distributed_music_positioning} as AOA fingerprints. In this work, we use MUSIC-based AOA estimates for localization. Unlike the channel power matrix, which capture both phase and distance information in a higher dimensional space, MUSIC extracts essential phase information alone, complementing the distance information captured by the RSS. Additionally, our previous research~\cite{my_wcl_letter} also justifies the effectiveness of combining RSS and MUSIC-based AOA estimates.

\vspace{\vertspacebeforesubsections}
\subsection{Decentralized Localization Methods in MIMO Systems}
Most localization methods discussed so far are centralized, relying on a central node to aggregate data for training and inference.
Some studies, such as~\cite{distr_loc_lis,distr_loc_lis2,distributed_music_positioning}, explore distributing localization to edge APs to reduce the burden on the central unit and reduce internodal data exchange. However, these approaches still require data transmission to a central unit for final inference. Our approach is among the first to propose a fully decentralized framework, eliminating any information exchange between APs and the CPU, enabling low-latency operation. The UE, instead, directly handles fusion through minimal computations that run in effectively constant time.

\vspace{\vertspacebeforesubsections}
\subsection{Proposed Framework Within the ML Paradigm}
The proposed framework introduces a novel shift from traditional centralized methods by leveraging ideas from various ML approaches. Instead of relying on a single model trained at the CPU using aggregated AOA and RSS data, each AP independently trains its own GPR model based on local data, echoing the principles of distributed ML~\cite{distr_ml}. Each AP uses only its own RSS and AOA data, resembling the multi-view learning paradigm where models learn from distinct perspectives without cross-AP data sharing~\cite{multi_view_ml}. By incorporating edge artificial-intelligence (AI), inference is carried out at the APs, enabling faster, low-latency decision-making~\cite{edge_ai}. The UE then combines the independent position estimates from all APs, following principles from ensemble learning~\cite{ensemble_learning}. Therefore, the proposed algorithm represents a confluence of distributed ML, edge AI, multi-view and ensemble learning. 

\vspace{\vertspacebeforesections}
\section{System Model}
\label{system_model}
\begin{figure}
\centering
\captionsetup{justification=centering}
\includegraphics[width=3.2 in]{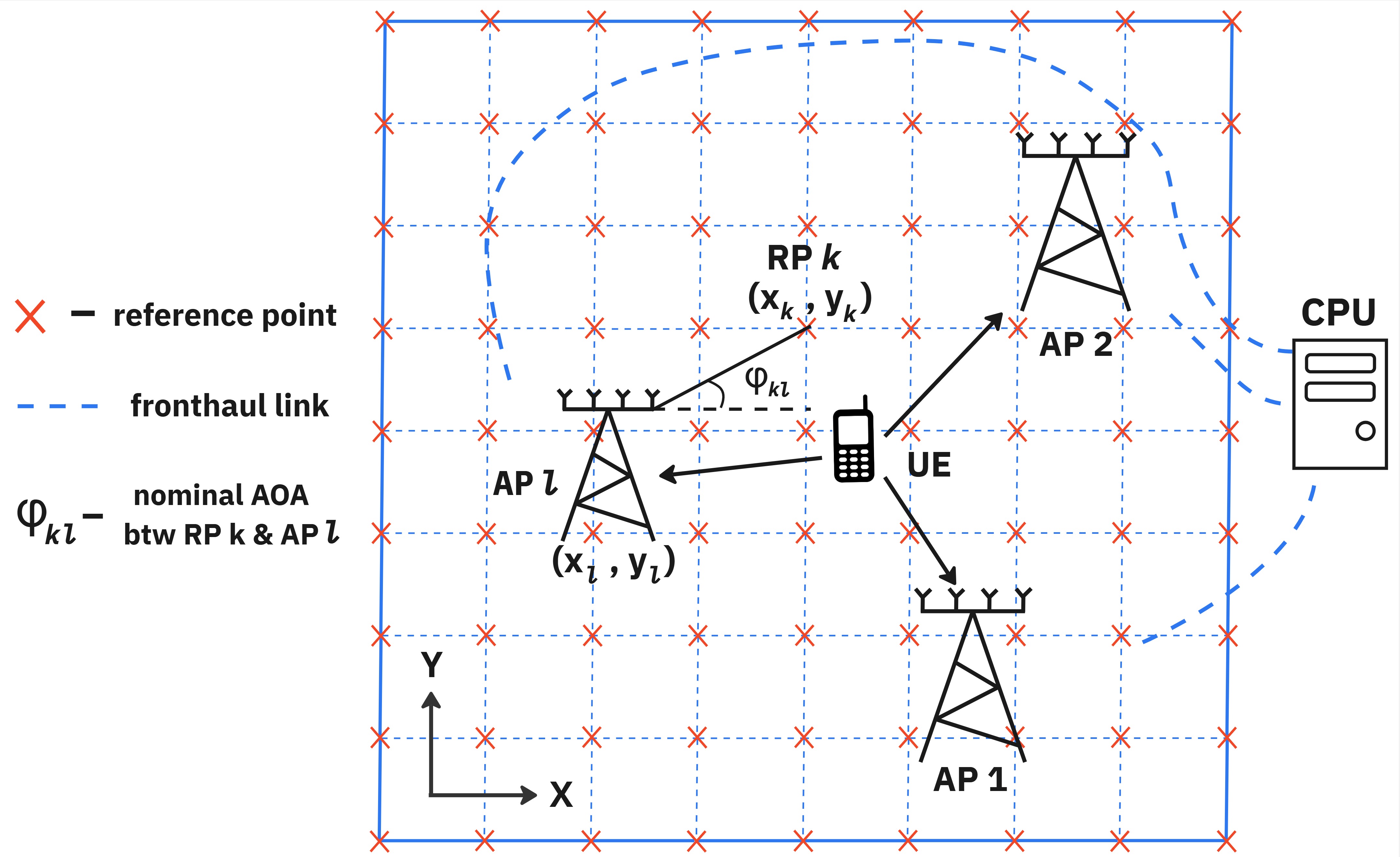}
\caption{System model illustrating the localization scenario.}
\label{system_model_fig}
\vspace{-2mm}
\end{figure}

A cell-free massive MIMO system consisting of $L$ geographically dispersed APs in the coverage area is considered for our localization scenario, as shown in Fig. \ref{system_model_fig}. 
Each of the $L$ APs is equipped with a uniform linear array (ULA) of $N$ omnidirectional antennas. As our focus is on two-dimensional (2D) localization, ULAs are sufficient to capture the angular information in the azimuth plane. The proposed approach can be readily extended to 3D localization scenarios by employing planar arrays. Without loss of generality, we assume that the ULA axis is oriented along the x-axis. The APs are connected to the CPU through fronthaul links. Although not directly involved in localization, the CPU coordinates the operation of the APs, as in other typical network operations within a cell-free massive MIMO system~\cite{cell_free_mimo_book}.

\begin{figure}
\centering
\captionsetup{justification=centering}
\includegraphics[width=2.8 in]{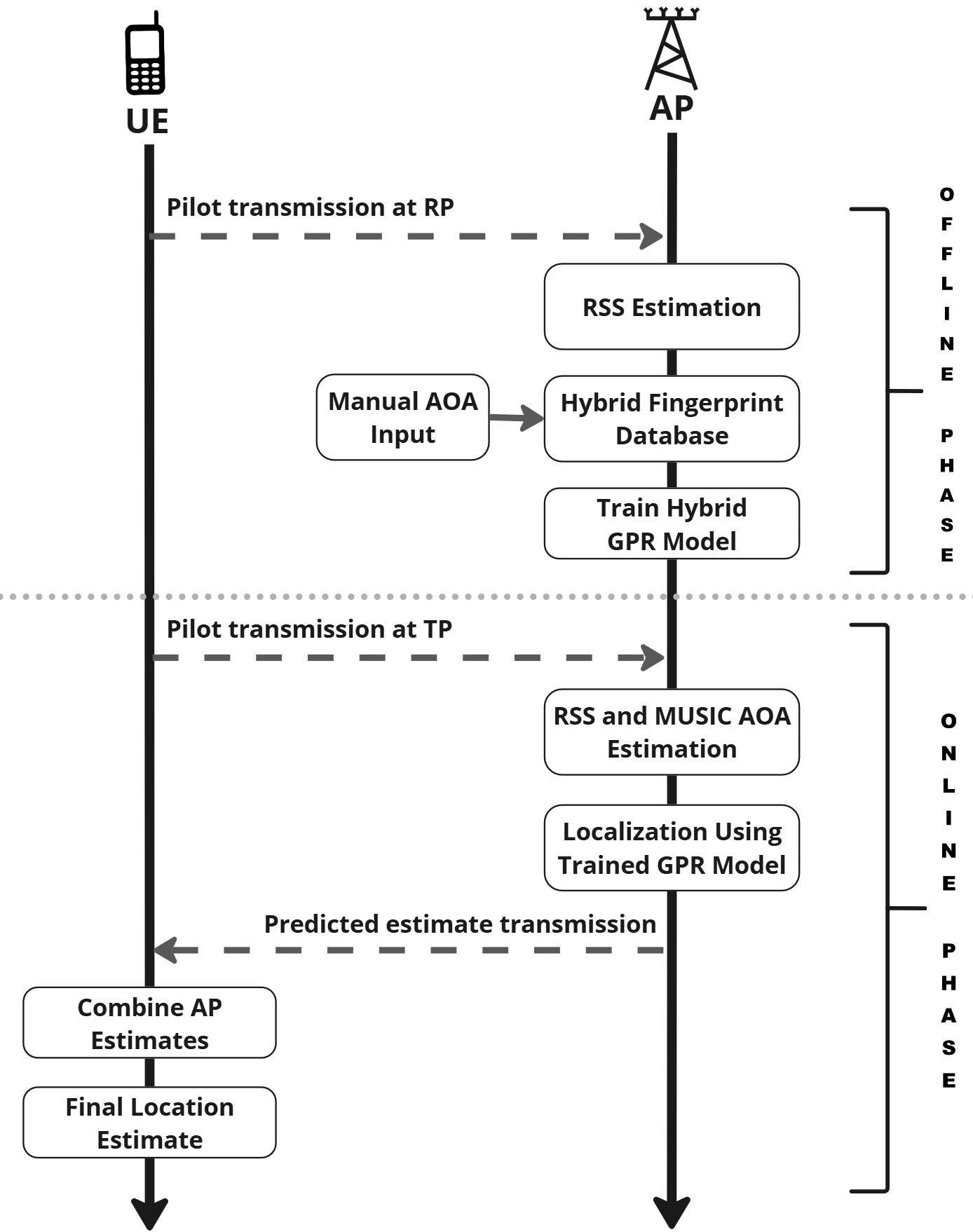}
\caption{Workflow diagram of the proposed distributed localization framework.}
\label{positioning_workflow}
\vspace{-2mm}
\end{figure}

The localization process is carried out in two distinct phases: the offline phase and the online phase. In the offline phase, a single-antenna UE is sequentially placed at predefined RPs, distributed throughout the system. RSS and AOA measurements from the UE placed at each of the $K$ RPs are collected from each AP and stored as a fingerprint database on the respective AP. Each AP trains a GPR model using its local RSS and AOA measurements as input features and the RP coordinates as target labels. By assuming a Gaussian process prior, the GPR model captures the relationship between inputs and outputs. In the online phase, each AP estimates the UE's location using its trained GPR model, based on RSS and AOA measurements from a test point. The APs transmit the predicted probabilistic position estimates to the UE, which then combines the $L$ AP estimates to determine the final predicted location, aiming to minimize the prediction error. The overall localization process is illustrated in Fig. \ref{positioning_workflow}.

The UE transmits a narrowband pilot vector $\boldsymbol{\psi} \in \mathbb{C}^{z \times 1}$, $\|\boldsymbol{\psi}\|^2 =z$ when placed at an RP during the offline phase or at a test point during the online phase. The pilot vector is assigned to the UE upon network access. Although a single-user case is considered in both offline and online phases for simplicity, the system model can be readily extended to support multi-user localization by assigning orthogonal pilot sequences to different UEs in the online phase. In addition, we assume that the UE executes coarse synchronization to correct for rough delay discrepancies in the channel, ensuring that the transmission of the pilot vector $\boldsymbol{\psi}$ is aligned with the network's timing requirements \cite{cell_free_mimo_book}. The signal transmitted by the UE reaches the APs after bouncing off the local scatterers surrounding the UE, which are modeled using a local scattering model \cite{mimo_book_correlation_matrix, disk_scatter_model}. The received signal $\mathbf{Y}_{k\ell} \in \mathbb{C}^{N \times z } $ at AP $\ell$ from the UE placed at RP $k$ is given by
\begin{equation}
\label{eqn1}
\textbf{Y}_{k\ell} = \sqrt{\rho}\textbf{h}_{k\ell}\boldsymbol{\psi}^H + \textbf{W}_{\ell},
\end{equation}
where $\rho$ is the transmit power of UE, $\textbf{h}_{k\ell} \in \mathbb{C}^{N \times 1 } $ is the narrowband channel from RP $k$ to AP $\ell$, and  $\textbf{W}_{\ell} \in \mathbb{C}^{N \times z}$ represents the additive white Gaussian noise~(AWGN) at AP $\ell$. The noise $\textbf{W}_{\ell}$ is assumed to be statistically independent of both the channel and the pilot vector. The elements of $\textbf{W}_{\ell}$ are independent and identically distributed (i.i.d) as $\mathcal{CN}(0,\sigma_n^2)$. The channel $\textbf{h}_{k\ell}$ is given by~\cite{joint_aoa_rss, sec_trans_positioning}
\begin{equation}
\label{eqn2}
\textbf{h}_{k\ell} = \sqrt{\frac{\beta_{k\ell}}{M}}\sum\limits_{m=1}^{M}\alpha_{k\ell}^m \textbf{a}(\theta^m_{k\ell}),
\end{equation}
where $M$ denotes the number of scattering paths, which can be arbitrarily large, $\alpha_{k\ell}^m \sim \mathcal{CN}(0,1)$ represents the small-scale fading  coefficient of the $m^{th}$ path, $\theta^m_{k\ell}$ is the azimuth AOA of the $m^{th}$ path and $\textbf{a}({\theta}^{m}_{k\ell}) \in \mathbb{C}^{N \times 1}$ is the array steering vector for a ULA. The steering vector is given by 
\begin{equation} 
\label{eqn3}
\resizebox{0.9\columnwidth}{!}{$
\textbf{a}(\theta^{m}_{k\ell}) = 
\begin{bmatrix}
1, e^{-j\frac{2\pi d}{\lambda}\cos(\theta^{m}_{k\ell})},\ \dots \ ,e^{-j\frac{2\pi (N-1) d}{\lambda}\cos(\theta^{m}_{k\ell})}
\end{bmatrix}^\top,
$}
\end{equation}
where $d$ is the antenna array spacing and $\lambda$ is the signal carrier wavelength. The angles ${\theta}^{m}_{k\ell}$ are i.i.d random variables, which are modeled as ${\theta}^{m}_{k\ell} = \varphi_{k\ell} + \Theta_m$. Here, $\varphi_{k\ell} = \mathbb{E}[{\theta}^{m}_{k\ell}]$ denotes the nominal azimuth AOA between the UE placed at RP $k$ and AP $\ell$, and corresponds to the azimuth of the LOS path connecting the RP and the ULA of the AP. The random perturbation $\Theta_m$ accounts for angular deviations due to scattering and is drawn independently for each path $m$ from an angular probability density function~(PDF) $f_{\Theta}({\theta})$.  This PDF is governed by the underlying scattering model and characterizes the statistical spread of arrival angles around the nominal azimuth AOA \cite{mimo_book_correlation_matrix}. For the large-scale fading coefficient $\beta_{k\ell}$, a log-distance path loss model is adopted, which is given by
\begin{equation}
\label{eqn4}
    \beta_{k\ell}[dB] = p^{0}_{\ell} - 10\gamma\log_{10}(d_{k\ell}/d_{\ell}^{0}) + \nu_{\scriptscriptstyle k\ell},
\end{equation}
where $d_{k\ell}$ is the three dimensional distance between RP $k$ and the center of the antenna array of AP $\ell$, $p^{0}_{\ell}$ is the path loss (in dB) at reference distance $d^{0}_{\ell}$, $\gamma$ is the path loss exponent,  and $\nu_{\scriptscriptstyle k\ell} \sim \mathcal{N}(0,\sigma_{\scriptscriptstyle SF}^2)$  is the shadowing noise. 

\vspace{\vertspacebeforesections}
\section{Fingerprint Extraction and Localization Strategy}
\label{fingerprint_extraction}
\subsection{RSS Fingerprint Extraction in the Offline Phase}
Upon receiving the signal $\textbf{Y}_{k\ell}$, the AP computes the RSS at AP $\ell$ from the UE placed at RP $k$. The RSS is mathematically expressed as $\xi_{k\ell} = \|\textbf{Y}_{k\ell}\|_F^2$. However, variations in the RSS arise due to small-scale fading in the received signal and the AWGN. To mitigate this, the RSS is calculated as the average value of the signal strength across a sufficiently large number of received symbols, i.e.,~\cite{ml_rss_surya, joint_aoa_rss}
\begin{equation} 
\label{eqn5}
\begin{split}
\widehat{\xi}_{k\ell} &\approx
 \mathbb{E}\{\|\textbf{Y}_{k\ell}\|_F^2\} = \mathbb{E}\{\|{\sqrt{\rho}\textbf{h}_{k\ell}}\boldsymbol{\psi}^H + \textbf{W}_{\ell}\|_F^2\}\\
&= \rho z\mathbb{E}\{\|\textbf{h}_{k\ell}\|^2\} + \mathbb{E}\{\|\textbf{W}_{\ell}\|_F^2\} \\
&= \rho zN\beta_{k\ell} + zN\sigma_n^2.
\end{split}
\end{equation} 
The approximation arises from the assumption that by averaging over enough samples, the RSS becomes less sensitive to small-scale fading and additive noise. This effect is a consequence of channel hardening, wherein the small-scale variability of the wireless channel diminishes as the number of independent channel realizations increases, in accordance with the law of large numbers~\cite{cell_free_mimo_book,joint_aoa_rss,ml_rss_surya}. Shadowing, however, persists as it is space-dependent and cannot be averaged out through temporal averaging~\cite{ml_rss_surya,fingerprint_savic}. 
For each AP $\ell$, define the RSS vector over $K$ RPs as $\boldsymbol{\xi}_{\ell}~=~[\widehat{\xi}_{\scriptscriptstyle 1\ell}^{\scriptscriptstyle dB}, \widehat{\xi}_{\scriptscriptstyle 2\ell}^{\scriptscriptstyle dB}, \  \dots\  , \widehat{\xi}_{\scriptscriptstyle K\ell}^{\scriptscriptstyle dB}]^\top \in \mathbb{R}^{K\times{1}}$, where $\widehat{\xi}_{\scriptscriptstyle k\ell}^{\scriptscriptstyle dB} = 10\log_{10}(\widehat{\xi}_{k\ell}/\rho)$ represents the estimated RSS at AP $\ell$ from the UE at RP $k$ in decibels~(dB).
The RSS vector is locally stored in the AP's memory and utilized for training the ML model used in localization. 

\vspace{\vertspacebeforesubsections}
\subsection{AOA Fingerprint Extraction in the Offline Phase}
The nominal AOA can be determined through geometric measurements during the offline phase using the known locations
of the APs and RPs. For the UE at RP $k$ with coordinates $\boldsymbol{q}_k = (x_k, y_k) \in \mathbb{R}^{1\times2}$ and AP $\ell$  positioned at $(x_{\ell}, y_{\ell})$ as shown in Fig. \ref{system_model_fig}, the nominal azimuth AOA is calculated as $\varphi_{k\ell} = \mathrm{atan2}(y_k - y_\ell, x_k - x_\ell)$, where atan2 denotes the four-quadrant arctangent. The computed AOA vector $\boldsymbol{\Phi}_{\ell} = [{\varphi}_{\scriptscriptstyle 1\ell}^{\circ}, \varphi_{\scriptscriptstyle 2\ell}^{\circ}, \  \dots\  , \varphi_{\scriptscriptstyle K\ell}^{\circ}]^\top \in \mathbb{R}^{K\times{1}}$ along with the RP position matrix $\boldsymbol{Q}_{\scriptscriptstyle RP} = [\boldsymbol{q}_1^\top,\boldsymbol{q}_2^\top,\ \dots \ , \boldsymbol{q}_K^\top]^\top \in \mathbb{R}^{K\times{2}}$ are stored in the respective AP memory, where $\varphi_{k\ell}^{\circ}$ represents the AOA measured in degrees. 

\vspace{\vertspacebeforesubsections}
\subsection{AOA Estimation in the Online Phase}
While RSS estimation in the online phase follows the same procedure as described in the offline phase in (\ref{eqn5}), for estimating the AOA in the online phase, the AP employs the MUSIC algorithm~\cite{music_schmidt,music_with_scatter}. For a UE located at a test point $\boldsymbol{q}_{\scriptscriptstyle TP} = (x_{\scriptscriptstyle TP},y_{\scriptscriptstyle TP})  \in \mathbb{R}^{1\times2}$, the MUSIC algorithm starts by estimating the covariance matrix of the received signal $\textbf{R}_{\scriptscriptstyle TP,\ell} \in \mathbb{R}^{N \times N}$ at AP $\ell$, expressed as~\cite{mimo_book_correlation_matrix}
\begin{equation} 
\label{eqn6}
\resizebox{0.91\columnwidth}{!}{$
\begin{split}
\textbf{R}_{\scriptscriptstyle TP,\ell} &= \mathbb{E}\{\textbf{Y}_{\scriptscriptstyle TP,\ell}\textbf{Y}_{\scriptscriptstyle TP,\ell}^H\} = \rho z\mathbb{E}\{\textbf{h}_{\scriptscriptstyle TP,\ell}\textbf{h}^H_{\scriptscriptstyle TP,\ell}\} + \mathbb{E}\{\textbf{W}_{\ell}\textbf{W}^H_{\ell}\}\\
&= \rho z\beta_{\scriptscriptstyle TP,\ell}\mathbb{E}\{\textbf{a}({\theta}^m_{\scriptscriptstyle TP,\ell})\textbf{a}^H({\theta}^m_{\scriptscriptstyle TP,\ell})\} + z\sigma_n^2\textbf{I}_N,
\end{split}
$}
\end{equation}
with, the $(p,q){th}$ element of matrix $\textbf{R}_{\scriptscriptstyle TP,\ell}$ given by~\cite{mimo_book_correlation_matrix} 
\begin{equation}
\label{eqn7}
\resizebox{0.9\columnwidth}{!}{$
[\textbf{R}_{\scriptscriptstyle TP,\ell}]_{pq} = \rho z\beta_{\scriptscriptstyle TP,\ell}\int e^{-j\frac{2\pi d}{\lambda}(p-q)\cos(\varphi_{\scriptscriptstyle TP,\ell} + {\theta})}f_{\Theta}({\theta})d{\theta} + z\sigma_n^2\delta_{pq}
$}.
\end{equation}
Similar to the RSS, $\textbf{R}_{\scriptscriptstyle TP,\ell}$ is estimated by averaging over a large, but finite number of received symbols. The eigenvector $\textbf{u}_1 \in \mathbb{R}^{N \times 1}$ corresponding to the largest eigenvalue of $\textbf{R}_{\scriptscriptstyle TP,\ell}$ span the signal subspace $\textbf{u}_s = [\textbf{u}_1] \in \mathbb{R}^{N \times 1} $, while the eigenvectors $\textbf{u}_2, \textbf{u}_3, ... ,\textbf{u}_N \in \mathbb{R}^{N \times 1}$ corresponding to the $N-1$ smallest eigenvalues span the noise subspace $\textbf{U}_n = [\textbf{u}_2,\textbf{u}_3,\ \dots \ ,\textbf{u}_N] \in \mathbb{R}^{N \times (N-1)}$. The orthogonality of the noise subspace and the
signal subspace implies that the matrix product ${\textbf{a}^H({\theta})\textbf{U}_n\textbf{U}^H_n\textbf{a}({\theta})}$ attains its minimum when the variable $\theta$ is equal to the true AOA $\varphi_{\scriptscriptstyle TP,\ell}^{\circ}$. Hence, the MUSIC pseudospectrum function defined as \cite{180_degree_ambiguity}
\begin{equation} 
\label{eqn8}
\mathrsfso{P}_{\scriptscriptstyle MUSIC}(\theta) = \frac{1}{{\textbf{a}^H({\theta})\textbf{U}_n\textbf{U}^H_n\textbf{a}({\theta})}},
\end{equation}
produces a sharp peak when $\theta = {\varphi}_{\scriptscriptstyle TP,\ell}^{\circ}$. Accordingly, the estimated AOA $\widehat{\varphi}_{\scriptscriptstyle TP,\ell}^{\circ}$ of the UE is obtained as the angle corresponding to this peak in the MUSIC pseudospectrum. In the multi-user localization case, the pseudospectrum exhibits distinct peaks corresponding to each user's AOA, which can be distinguished using their orthogonal pilot sequences~\cite{class_and_mod_doa}.

\vspace{\vertspacebeforesubsections}
\subsection{Localization Strategy Overview}
Each AP independently computes its own AOA vector $\boldsymbol{\Phi}_{\ell}$ and RSS vector $\boldsymbol{\xi}_{\ell}$ and remains unaware of the AOA and RSS vectors of other APs. Using the computed AOA and RSS vectors, AP $\ell$ constructs a combined hybrid fingerprint matrix $\boldsymbol{\Upsilon}_{\ell} = [\boldsymbol{\xi}_{\ell},\boldsymbol{\Phi}_{\ell}]\in \mathbb{R}^{K \times 2}$. 
The combined hybrid fingerprint matrix $\boldsymbol{\Upsilon}_{\ell}$ and the RP position matrix $\boldsymbol{Q}_{\scriptscriptstyle RP}$ collectively form the fingerprint database of AP $\ell$ and is stored in its local memory. This database is used by the AP to learn a GPR model for each RP coordinate vector in the offline phase, i.e.,
\begin{equation} 
\label{eqn9}
\begin{split}
&col_1(\boldsymbol{Q}_{\scriptscriptstyle RP}) = [x_1,x_2,\  \dots\ ,x_K]^\top = f_{1\ell}(\boldsymbol{\Upsilon}_{\ell}) + \boldsymbol{\epsilon}_{\scriptscriptstyle 1\ell} \\
&col_2(\boldsymbol{Q}_{\scriptscriptstyle RP}) = [y_1,y_2,\  \dots\ ,y_K]^\top = f_{2\ell}(\boldsymbol{\Upsilon}_{\ell}) + \boldsymbol{\epsilon}_{\scriptscriptstyle 2\ell}
\end{split}
\end{equation}
for $k = \{1,2,\ \dots \ ,K\}$ and $\ell = \{1,2,\ \dots \ ,L\}$. The vectors $\boldsymbol{\epsilon}_{\scriptscriptstyle 1\ell},  \boldsymbol{\epsilon}_{\scriptscriptstyle 2\ell} \in \mathbb{R}^{K \times 1}$ represent observation noise arising from measurement inaccuracies inherent in practical implementations of the fingerprinting process. Each component of $\boldsymbol{\epsilon}_{\scriptscriptstyle 1\ell},  \boldsymbol{\epsilon}_{\scriptscriptstyle 2\ell}$ is modeled as a zero-mean Gaussian random variable with variances $\sigma_{\epsilon_{\scriptscriptstyle 1\ell}}^2$ and $\sigma_{\epsilon_{\scriptscriptstyle 2\ell}}^2$, respectively.

Using the learned functions $f_{1\ell}(\cdot)$ and $f_{2\ell}(\cdot)$ from the GPR model, AP $\ell$ estimates the test point coordinate $col_i(\boldsymbol{q}_{\scriptscriptstyle TP})$ as a Gaussian random variable with mean $\mu_{\scriptscriptstyle i\ell}$ and variance ${v}_{\scriptscriptstyle i\ell}$, $i\in\{1,2\}, \ell \in \{1,2,\ \dots \ ,L\}$, using $\widehat{\xi}_{\scriptscriptstyle TP,\ell}^{\scriptscriptstyle dB}$ and $\widehat{\varphi}_{\scriptscriptstyle TP,\ell}^{\circ}$ in the online phase. Each AP transmits its local position estimate, comprising the mean and variance for each coordinate, to the UE over the downlink. The UE then aggregates the mean and variance estimates from $L$ APs based on the chosen variant of the distributed localization framework, yielding a final position coordinate estimate for the test point, which is also a Gaussian random variable, with mean $\mu_{i}$ and variance ${v}_{i}$. Details of these variants are further elaborated in Section \ref{distributed_loc_framework}. The mean vector $\overline{\boldsymbol{q}}_{est} = (\mu_1,\mu_2) =(\overline{x}_{\scriptscriptstyle est},\overline{y}_{\scriptscriptstyle est}) \in \mathbb{R}^{1 \times 2}$ is the final predicted position of $\boldsymbol{q}_{\scriptscriptstyle TP}$, while the variances ${v}_{1}$ and ${v}_{2}$ model the uncertainties in the estimates of $x_{\scriptscriptstyle TP}$ and $y_{\scriptscriptstyle TP}$ respectively. 

The two main objectives of the localization process are:
\begin{enumerate}
    \item To minimize the localization error by bringing 
$\overline{\boldsymbol{q}}_{est}$ as close as possible to $\boldsymbol{q}_{\scriptscriptstyle TP}$.
    \item To reduce the variances of the estimates ${v}_{i}$, effectively decreasing the uncertainty around the prediction.
\end{enumerate}

\vspace{\vertspacebeforesections}
\section{Distributed Localization Framework}
\label{distributed_loc_framework}
\subsection{GPR-Based Fingerprint Localization}
\label{GPR_based_loc}
Equations (\ref{eqn4}) and (\ref{eqn5}) show that the RSS (in dB) of a UE measured at an AP is inversely proportional to the logarithm of the distance between them. The  distance information (from $\boldsymbol{\xi}_{\ell}$) and the angle information (from $\boldsymbol{\Phi}_{\ell}$) derived from the hybrid fingerprint $\boldsymbol{\Upsilon}_{\ell}$, serve as fundamental elements for the GPR model to estimate $\boldsymbol{q}_{\scriptscriptstyle TP}$ using the hybrid test point vector $\boldsymbol{\Upsilon}_{\ell}^{\scriptscriptstyle TP} = [\widehat{\xi}_{\scriptscriptstyle TP,\ell}^{\scriptscriptstyle dB}$, $\widehat{\varphi}_{\scriptscriptstyle TP,\ell}^{\circ}] \in \mathbb{R}^{1 \times 2}$ in the online phase.

The GPR approach assumes that each of the $2L$ functions $f_{1\ell}(.)$  and $f_{2\ell}(.)$ follow Gaussian processes with zero mean and user-defined covariance functions $\textit{\textbf{C}}_{1\ell}$ and $\textit{\textbf{C}}_{2\ell}$, respectively, for $\ell = \{1,2,\ \dots \ ,L\}$ \cite{fingerprint_savic,gpr_book_mit}. To model the covariance functions, we adopt the squared exponential kernel, i.e.,
\begin{equation} 
\label{eqn10}
{k}_{\scriptscriptstyle i\ell}(\boldsymbol{r},\boldsymbol{r'}) = b_{\scriptscriptstyle i\ell}^2\exp\left(-\frac{\|\boldsymbol{r}-\boldsymbol{r'}\|^2}{2\varrho_{\scriptscriptstyle i\ell}}\right).
\end{equation}
For GPR modeling, the vectors $\boldsymbol{r},\boldsymbol{r}'$ $\in \{\boldsymbol{\Upsilon}_{j\ell},\boldsymbol{\Upsilon}_{\scriptscriptstyle TP}\}$. Here $\boldsymbol{\Upsilon}_{j\ell} = row_j(\boldsymbol{\Upsilon}_{\ell})$ and $b_{\scriptscriptstyle i\ell}^2, \varrho_{\scriptscriptstyle i\ell}$ are hyperparameters, for~$i \in~\{ 1,2 \}$. While we use the squared exponential kernel in this work, the proposed framework remains compatible with any valid choice of kernel function~\cite{gpr_book_mit}. Eq. (\ref{eqn9}) can be recast using the Gaussian process assumption of functions $f_{i\ell}(.)$ as 
\begin{equation} 
    col_i(\boldsymbol{Q}_{\scriptscriptstyle RP}) \sim \\ \mathcal{N}\left(\textbf{0}_{K\times1},\bm{\mathcal{K}}_{\scriptscriptstyle i\ell}(\boldsymbol{\Upsilon}_{\ell},\boldsymbol{\Upsilon}_{\ell}) + \sigma_{\epsilon_{\scalebox{0.5}{$i\ell$}}}^2\textbf{I}_K \right )
\label{eqn11}
\end{equation} 
for $i$ $\in$ $\{1,2\}$. 
Here,
\begin{align*}   
&\textit{\textbf{C}}_{i\ell} = 
\bm{\mathcal{K}}_{\scriptscriptstyle i\ell}(\boldsymbol{\Upsilon}_{\ell},\boldsymbol{\Upsilon}_{\ell}) \\
&= \resizebox{0.86\columnwidth}{!}{$
\begin{bmatrix} 
   {k}_{\scriptscriptstyle i\ell}(\boldsymbol{\Upsilon}_{1\ell},\boldsymbol{\Upsilon}_{1\ell}) & {k}_{\scriptscriptstyle i\ell}(\boldsymbol{\Upsilon}_{1\ell},\boldsymbol{\Upsilon}_{2\ell}) & \dots &
    {k}_{\scriptscriptstyle i\ell}(\boldsymbol{\Upsilon}_{1\ell},\boldsymbol{\Upsilon}_{K\ell}) \\
    {k}_{\scriptscriptstyle i\ell}(\boldsymbol{\Upsilon}_{2\ell},\boldsymbol{\Upsilon}_{1\ell}) &
    {k}_{\scriptscriptstyle i\ell}(\boldsymbol{\Upsilon}_{2\ell},\boldsymbol{\Upsilon}_{2\ell}) & \dots &
    {k}_{\scriptscriptstyle i\ell}(\boldsymbol{\Upsilon}_{2\ell},\boldsymbol{\Upsilon}_{K\ell}) \\
    \vdots & \vdots &\ddots & \vdots\\
    {k}_{\scriptscriptstyle i\ell}(\boldsymbol{\Upsilon}_{K\ell},\boldsymbol{\Upsilon}_{1\ell}) &
    {k}_{\scriptscriptstyle i\ell}(\boldsymbol{\Upsilon}_{K\ell},\boldsymbol{\Upsilon}_{2\ell}) & \dots &
    {k}_{\scriptscriptstyle i\ell}(\boldsymbol{\Upsilon}_{K\ell},\boldsymbol{\Upsilon}_{K\ell}) \\    
\end{bmatrix}
$}_{K\times K} 
\tag{12}
\label{eqn12}
\end{align*}
for $i = \{1,2\}$, models the covariance between all pairs of training fingerprint vectors of AP $\ell$.

The training process at AP $\ell$ involves learning the hyperparameters $b_{\scriptscriptstyle i\ell}^2, \varrho_{\scriptscriptstyle i\ell}$ and the error variances $\sigma_{\epsilon_{\scriptscriptstyle i\ell}}^2$ from its fingerprint matrix $\boldsymbol{\Upsilon}_{\ell}$ using maximum-likelihood approach as~\cite{ml_rss_surya,gpr_book_mit}
\begin{equation} 
\tag{13}
\label{eqn13}
\overline{\boldsymbol{\Gamma}}_{\scriptscriptstyle i\ell} = \underset{\boldsymbol{\Gamma}_{\scriptscriptstyle i\ell}}{\operatorname{arg\,max}} \Big\{ \log\big(p\left(\text{col}_i(\boldsymbol{Q}_{\scriptscriptstyle RP}) \mid \boldsymbol{\Upsilon}_{\ell}, \boldsymbol{\Gamma}_{\scriptscriptstyle i\ell}\right)\big) \Big\}, 
\end{equation}
where $\boldsymbol{\Gamma}_{\scriptscriptstyle i\ell} =[b_{\scriptscriptstyle i\ell}^2, \varrho_{\scriptscriptstyle i\ell}$,  $\sigma_{\epsilon_{\scriptscriptstyle i\ell}}^2]^T \in \mathbb{R}^{3 \times 1}$ and $\overline{\boldsymbol{\Gamma}}_{\scriptscriptstyle i\ell}$ represents the optimized vector after learning. The non-convex optimization problem in (\ref{eqn13}) can be solved using gradient ascent methods, such as stochastic gradient or conjugate gradient~\cite{gpr_book_mit,ml_rss_surya}.

The joint Gaussian distributions of the RP position coordinates with the TP position coordinates is given by
\begin{align*}
& \begin{bmatrix}
    col_i(\boldsymbol{Q}_{\scriptscriptstyle RP}) \\
    col_i(\boldsymbol{q}_{\scriptscriptstyle TP})
\end{bmatrix}
\sim \\
&\resizebox{0.89\columnwidth}{!}{$
\mathcal{N}
\left(
\textbf{0}_{(K+1)\times1},
\begin{bmatrix}
    \bm{\mathcal{K}}_{\scriptscriptstyle i\ell}(\boldsymbol{\Upsilon}_{\ell},\boldsymbol{\Upsilon}_{\ell}) + \sigma_{\epsilon_{\scriptscriptstyle i\ell}}^2\textbf{I}_K & 
    \bm{\mathcal{K}}_{\scriptscriptstyle i\ell}(\boldsymbol{\Upsilon}_{\ell},\boldsymbol{\Upsilon}_{\ell}^{\scriptscriptstyle TP})
    \\
    \bm{\mathcal{K}}_{\scriptscriptstyle i\ell}(\boldsymbol{\Upsilon}_{\ell}^{\scriptscriptstyle TP},\boldsymbol{\Upsilon}_{\ell})
    &
    {k}_{\scriptscriptstyle i\ell}(\boldsymbol{\Upsilon}_{\ell}^{\scriptscriptstyle TP},\boldsymbol{\Upsilon}_{\ell}^{\scriptscriptstyle TP})
\end{bmatrix}
\right )
$}
\tag{14}
\label{eqn14}
\end{align*}
for $i$ $\in$ $\{1,2\}$. Here,
\begin{align*}   
&\bm{\mathcal{K}}_{\scriptscriptstyle i\ell}^\top(\boldsymbol{\Upsilon}_{\ell}^{\scriptscriptstyle TP},\boldsymbol{\Upsilon}_{\ell}) = \bm{\mathcal{K}}_{\scriptscriptstyle i\ell}(\boldsymbol{\Upsilon}_{\ell},\boldsymbol{\Upsilon}_{\ell}^{\scriptscriptstyle TP}) \\ 
&= \resizebox{0.85\columnwidth}{!}{$
\begin{bmatrix} 
    {k}_{\scriptscriptstyle i\ell}(\boldsymbol{\Upsilon}_{1\ell},\boldsymbol{\Upsilon}_{\ell}^{\scriptscriptstyle TP}) &{k}_{\scriptscriptstyle i\ell}(\boldsymbol{\Upsilon}_{2\ell},\boldsymbol{\Upsilon}_{\ell}^{\scriptscriptstyle TP}) & \cdots &
    {k}_{\scriptscriptstyle i\ell}(\boldsymbol{\Upsilon}_{K\ell},\boldsymbol{\Upsilon}_{\ell}^{\scriptscriptstyle TP})  
\end{bmatrix}
$}_{K\times 1}
\tag{15}
\label{eqn15}
\end{align*}
for $i = \{1,2\}$, models the covariance between the $K$ training fingerprint vectors and the test vector of AP $\ell$. Conditioning the joint distribution in (\ref{eqn14}) on the fingerprint database matrices $\boldsymbol{\Upsilon}_\ell$ and $\boldsymbol{Q}_{\scriptscriptstyle RP}$ and the test vector $\boldsymbol{\Upsilon}_{\ell}^{\scriptscriptstyle TP}$, we obtain normal posterior densities for $\boldsymbol{q}_{\scriptscriptstyle TP}$, i.e.,
\begin{equation} 
\resizebox{0.89\columnwidth}{!}{$
p\left(col_i(\boldsymbol{q}_{\scriptscriptstyle TP})|col_i(\boldsymbol{Q}_{\scriptscriptstyle RP}),\boldsymbol{\Upsilon}_\ell,\boldsymbol{\Upsilon}_{\ell}^{\scriptscriptstyle TP}\right) ~  \sim \mathcal{N}(\mu{\scriptscriptstyle i\ell},{v}_{\scriptscriptstyle i\ell})
$}
\tag{16}
\label{eqn16}
\end{equation}
with the mean and variance 
\begin{align*}  
&\mu_{\scriptscriptstyle i\ell}  =  \bm{\mathcal{K}}_{\scriptscriptstyle i\ell}(\boldsymbol{\Upsilon}_{\ell}^{\scriptscriptstyle TP},\boldsymbol{\Upsilon}_{\ell})[\bm{\mathcal{K}}_{\scriptscriptstyle i\ell}(\boldsymbol{\Upsilon}_{\ell},\boldsymbol{\Upsilon}_{\ell}) + \sigma_{\epsilon_{\scalebox{0.5}{$i\ell$}}}^2\textbf{I}_K]^{-1}col_i(\boldsymbol{p}_{\scriptscriptstyle RP}) \\
\tag{17}
\label{eqn17}
&{v}_{\scriptscriptstyle i\ell} = {k}_{\scriptscriptstyle i\ell}(\boldsymbol{\Upsilon}_{\ell}^{\scriptscriptstyle TP},\boldsymbol{\Upsilon}_{\ell}^{\scriptscriptstyle TP})\  \\
&\quad -  \bm{\mathcal{K}}_{\scriptscriptstyle i\ell}(\boldsymbol{\Upsilon}_{\ell}^{\scriptscriptstyle TP},\boldsymbol{\Upsilon}_{\ell})[\bm{\mathcal{K}}_{\scriptscriptstyle i\ell}(\boldsymbol{\Upsilon}_{\ell},\boldsymbol{\Upsilon}_{\ell}) + \sigma_{\epsilon_{\scalebox{0.5}{$i\ell$}}}^2\textbf{I}_K]^{-1}\bm{\mathcal{K}}_{\scriptscriptstyle i\ell}(\boldsymbol{\Upsilon}_{\ell},\boldsymbol{\Upsilon}_{\ell}^{\scriptscriptstyle TP})
\end{align*}
respectively. For each AP $\ell$, the mean vector $\overline{\boldsymbol{q}}^{est}_\ell = (\mu_{\scriptscriptstyle 1\ell},\mu_{\scriptscriptstyle 2\ell}) = (\overline{x}_\ell^{est},\overline{y}_\ell^{est})$ serves as the minimum mean square error (MMSE) estimate of $\boldsymbol{q}_{\scriptscriptstyle TP}$, while the variances ${v}_{\scriptscriptstyle 1\ell}$ and ${v}_{\scriptscriptstyle 2\ell}$ model the uncertainty in the estimate of x and y coordinates respectively, for AP $\ell$. These estimates are transmitted to the UE in the downlink.

\vspace{\vertspacebeforesubsections}
\subsection{Aggregation of Position Estimates by the UE}
\label{UE_aggregation}
Once the estimates from the $L$ APs are available, the UE combines these estimates to derive a single estimate using one of the following algorithmic variants: distributed-median, distributed-mean, distributed-Bayesian, or distributed-z-score. These algorithms define distinct strategies for aggregating the GPR-based predictions from the $L$ APs by the UE, forming the core of the distributed localization framework. Each variant is designed to offer unique advantages, enabling optimized performance under varying conditions and localization scenarios.

\subsubsection{Distributed-Median Algorithm}
In this variant, the median of the $L$ MMSE estimates derived from the GPR outputs of $L$ APs, is computed independently for each coordinate dimension and selected as the final system estimate. The rationale is that, under the log-normal shadowing model, the median of the estimated coordinates corresponds to the estimate from the AP experiencing the least shadowing, thus providing the most accurate RSS and AOA estimates at the test point. This approach leverages the resistance of the median to outliers, ensuring that the final position estimate is more accurate and reliable in the presence of variable shadowing effects. The procedure is detailed in Algorithm \ref{algo:median_coordinates} for clarity.
\begin{algorithm}[h]
  \caption{Distributed-Median Algorithm}
  \label{algo:median_coordinates}
  \textbf{Input:} List of coordinate pairs \( \overline{\boldsymbol{q}}^{est}_\ell = (\overline{x}_\ell^{est}, \overline{y}_\ell^{est}) \) for \( \ell = \{1, 2, \ldots, L\} \)\\
  \textbf{Output:} Median coordinates \( \overline{\boldsymbol{q}}_{\scriptscriptstyle DD} = (\overline{x}_{\scriptscriptstyle DD}, \overline{y}_{\scriptscriptstyle DD}) \)
  
  \begin{algorithmic}[1]
    \State Sort the list of \( x \)-coordinates $\boldsymbol{x}_{\scriptscriptstyle D}$ = \( [\overline{x}_1^{est}, \overline{x}_2^{est}, \ldots, \overline{x}_L^{est}] \in \mathbb{R}^{1 \times L}  \)  and denote the sorted list as $\boldsymbol{x}_{\scriptscriptstyle sort}$ = \([\overline{x}_1^{\scriptscriptstyle sort},  \overline{x}_2^{\scriptscriptstyle sort},  \ldots, \overline{x}_L^{\scriptscriptstyle sort}] \in \mathbb{R}^{1 \times L} \)
    \State Sort the list of \( y \)-coordinates  $\boldsymbol{y}_{\scriptscriptstyle D}$ = \( [\overline{y}_1^{est}, \overline{y}_2^{est}, \ldots, \overline{y}_L^{est}] \in \mathbb{R}^{1 \times L} \) and denote the sorted list as $\boldsymbol{y}_{\scriptscriptstyle sort}$ = \( [\overline{y}_1^{\scriptscriptstyle sort}, \overline{y}_2^{\scriptscriptstyle sort}, \ldots, \overline{y}_L^{\scriptscriptstyle sort}] \in \mathbb{R}^{1 \times L} \)
    
    \If{L is odd}
      \State \( \overline{x}_{\scriptscriptstyle DD} \gets \overline{x}_{\scalebox{0.5}{${(L+1)}/{2}$}}^{\scriptscriptstyle sort} \) \Comment{Median \( x \)-coordinate}
      \State \( \ell_{\scriptscriptstyle x} \gets \) index of \( \overline{x}_{\scalebox{0.5}{${(L+1)}/{2}$}}^{\scriptscriptstyle sort} \)  in $\boldsymbol{x}_{\scriptscriptstyle D}$
      \State \( \overline{y}_{\scriptscriptstyle DD} \gets \overline{y}_{\scalebox{0.5}{${(L+1)}/{2}$}}^{\scriptscriptstyle sort} \) \Comment{Median \( y \)-coordinate}
      \State \( \ell_{\scriptscriptstyle y} \gets \) index of \( \overline{y}_{\scalebox{0.5}{${(L+1)}/{2}$}}^{\scriptscriptstyle sort} \) in $\boldsymbol{y}_{\scriptscriptstyle D}$
    \Else
      \State \( \overline{x}_{\scriptscriptstyle DD} \gets \frac{\overline{x}_{L/2}^{\scriptscriptstyle sort} + \overline{x}_{L/2+1}^{\scriptscriptstyle sort}}{2} \) \Comment{Median \( x \)-coordinate}
      \State \( \ell_{\scriptscriptstyle x\scalebox{0.4}{1}} \gets \) index of \( \overline{x}_{\scriptscriptstyle L/2}^{\scriptscriptstyle sort}  \) in $\boldsymbol{x}_{\scriptscriptstyle D}$
      \State \( \ell_{\scriptscriptstyle x\scalebox{0.4}{2}} \gets \) index of \( \overline{x}_{\scriptscriptstyle L/2+1}^{\scriptscriptstyle sort}  \) in $\boldsymbol{x}_{\scriptscriptstyle D}$
      \State \( \overline{y}_{\scriptscriptstyle DD} \gets \frac{\overline{y}_{L/2}^{\scriptscriptstyle sort} + \overline{y}_{L/2+1}^{\scriptscriptstyle sort}}{2} \)  \Comment{Median \( y \)-coordinate}
      \State \( \ell_{\scriptscriptstyle y\scalebox{0.4}{1}} \gets \) index of \( \overline{y}_{\scriptscriptstyle L/2}^{\scriptscriptstyle sort}  \) in $\boldsymbol{y}_{\scriptscriptstyle D}$
      \State \( \ell_{\scriptscriptstyle y\scalebox{0.4}{2}} \gets \) index of \( \overline{y}_{\scriptscriptstyle L/2+1}^{\scriptscriptstyle sort}  \) in $\boldsymbol{y}_{\scriptscriptstyle D}$
    \EndIf
    
    \State Return \( \overline{\boldsymbol{q}}_{\scriptscriptstyle DD} \gets (\overline{x}_{\scriptscriptstyle DD}, \overline{y}_{\scriptscriptstyle DD})\) 
    \end{algorithmic}
\end{algorithm}

The median coordinate vector $\overline{\boldsymbol{q}}_{\scriptscriptstyle DD} = \overline{\boldsymbol{q}}_{est} = (\overline{x}_{\scriptscriptstyle DD}, \overline{y}_{\scriptscriptstyle DD}) \in \mathbb{R}^{1 \times 2}$ is final predicted position. The variance of the position estimate is denoted as $\boldsymbol{v}_{\scriptscriptstyle DD} =  (v^{\scriptscriptstyle DD}_1,v^{\scriptscriptstyle DD}_2) = ({v}_{1},{v}_{2}) \in \mathbb{R}^{1 \times 2}$ and is equal to $(v_{ \scriptscriptstyle 1\ell_{x}},v_{ \scriptscriptstyle 1\ell_{y}})$ when $L$ is odd, and equal to  $\left( 
  \frac{
    \displaystyle v_{\scriptscriptstyle 1\ell_{x\scalebox{0.4}{1}}} + 
    v_{\scriptscriptstyle 1\ell_{y\scalebox{0.4}{1}}}
  }{4},\; 
  \frac{
    \displaystyle v_{\scriptscriptstyle 2\ell_{x\scalebox{0.4}{2}}} + 
    v_{\scriptscriptstyle 2\ell_{x\scalebox{0.4}{2}}}
  }{4}
\right)$ when $L$ is even (see Appendix \ref{appendixa}). This variance will be utilized to assess the uncertainty associated with the position estimates.

\subsubsection{Distributed-Mean Algorithm}
The variance of the distributed-median GPR estimate is equal to the variance of a selected AP's GPR model (when $L$ is odd), or at best, one-fourth the sum of variances from two APs (when $L$ is even). The median approach does not leverage the variances of estimates from all APs to reduce localization uncertainty. This motivates the distributed-mean GPR approach, where the mean of the $L$  MMSE estimates derived from the GPR outputs of the $L$ APs is independently computed for each coordinate dimension and selected as the final system estimate. Thus, the final position estimate in the distributed-mean approach is given by  $\overline{\boldsymbol{q}}_{\scriptscriptstyle DM} = \overline{\boldsymbol{q}}_{est} = (\overline{x}_{\scriptscriptstyle DM}, \overline{y}_{\scriptscriptstyle DM}) = \left(\sum\limits_{{\ell=1}}^{L}{\dfrac{\overline{x}_\ell^{est}}{\scriptstyle L}},\sum\limits_{{\ell=1}}^{L}{\dfrac{\overline{y}_\ell^{est}}{\scriptstyle L}}\right)
 \in \mathbb{R}^{1 \times 2}$. 

The variance of the final position estimate is denoted as $\boldsymbol{v}_{\scriptscriptstyle DM} = (v^{\scriptscriptstyle DM}_1,v^{\scriptscriptstyle DM}_2) = ({v}_{1},{v}_{2}) \in \mathbb{R}^{1 \times 2}$ and is equal to $\left(\sum\limits_{{\ell=1}}^{L}{\dfrac{v_{\scriptscriptstyle 1\ell}}{\scriptstyle L^2}},\sum\limits_{{\ell=1}}^{L}{\dfrac{v_{\scriptscriptstyle 2\ell}}{\scriptstyle L^2}}\right)$. The variance of each coordinate is reduced by a factor of $L$ compared to the average variance of the AP predictions, resulting in lower overall uncertainty and improved estimate quality compared to the distributed-median approach. (see Appendix \ref{appendixa}). 

\subsubsection{Distributed-Bayesian Algorithm}
\label{disributed_bayesian}
Here, we attempt to derive a joint position estimate by leveraging Bayesian inference, with the goal of further reducing position estimate variance compared to the distributed-mean approach. Ideally the UE would like to compute $p\left(col_i(\boldsymbol{q}_{\scriptscriptstyle TP}) | col_i(\boldsymbol{Q}_{\scriptscriptstyle RP}),\boldsymbol{\Upsilon}_1,\boldsymbol{\Upsilon}_{1}^{\scriptscriptstyle TP},\boldsymbol{\Upsilon}_2,\boldsymbol{\Upsilon}_{2}^{\scriptscriptstyle TP},\dots, \boldsymbol{\Upsilon}_L,\boldsymbol{\Upsilon}_{L}^{\scriptscriptstyle TP}\right)$. Using Bayes' rule we can write this as, 
\begin{equation}
\tag{18}
\label{eqn18}
\resizebox{\columnwidth}{!}{$
\begin{aligned}
&p\Big(col_i(\boldsymbol{q}_{\scriptscriptstyle TP}) \;\big|\; col_i(\boldsymbol{Q}_{\scriptscriptstyle RP}),\, \boldsymbol{\Upsilon}_1,\, \boldsymbol{\Upsilon}_{1}^{\scriptscriptstyle TP},\, \dots,\, \boldsymbol{\Upsilon}_L,\, \boldsymbol{\Upsilon}_{L}^{\scriptscriptstyle TP} \Big) = \\
&\frac{ p\Big(col_i(\boldsymbol{Q}_{\scriptscriptstyle RP}),\, \boldsymbol{\Upsilon}_1,\, \boldsymbol{\Upsilon}_{1}^{\scriptscriptstyle TP},\, \dots,\, \boldsymbol{\Upsilon}_L,\, \boldsymbol{\Upsilon}_{L}^{\scriptscriptstyle TP} \;\big|\; col_i(\boldsymbol{q}_{\scriptscriptstyle TP})\Big) \; p\Big(col_i(\boldsymbol{q}_{\scriptscriptstyle TP})\Big) }{ p\Big(col_i(\boldsymbol{Q}_{\scriptscriptstyle RP}),\, \boldsymbol{\Upsilon}_1,\, \boldsymbol{\Upsilon}_{1}^{\scriptscriptstyle TP},\, \dots,\, \boldsymbol{\Upsilon}_L,\, \boldsymbol{\Upsilon}_{L}^{\scriptscriptstyle TP} \Big) }.
\end{aligned}
$}
\end{equation}
In our localization model, we assume that the measurements from different APs are conditionally independent given the user's position. This allows us to express the joint likelihood of observations as
\begin{equation}
\tag{19}
\label{eqn19}
\resizebox{0.87\columnwidth}{!}{$
\begin{aligned}
&p\Big(col_i(\boldsymbol{Q}_{\scriptscriptstyle RP}),\, \boldsymbol{\Upsilon}_1,\, \boldsymbol{\Upsilon}_{1}^{\scriptscriptstyle TP},\, \dots,\, \boldsymbol{\Upsilon}_L,\, \boldsymbol{\Upsilon}_{L}^{\scriptscriptstyle TP} \;\big|\; col_i(\boldsymbol{q}_{\scriptscriptstyle TP})\Big) \\
&= p\Big(col_i(\boldsymbol{Q}_{\scriptscriptstyle RP}),\, \boldsymbol{\Upsilon}_1,\, \boldsymbol{\Upsilon}_{1}^{\scriptscriptstyle TP} \;\big|\; col_i(\boldsymbol{q}_{\scriptscriptstyle TP})\Big) \\
&\; \times \; 
p\Big(col_i(\boldsymbol{Q}_{\scriptscriptstyle RP}),\, \boldsymbol{\Upsilon}_2,\, \boldsymbol{\Upsilon}_{2}^{\scriptscriptstyle TP} \;\big|\; col_i(\boldsymbol{q}_{\scriptscriptstyle TP})\Big) \; \times \\
&\hspace{0.5cm} \dots \; \times \; p\Big(col_i(\boldsymbol{Q}_{\scriptscriptstyle RP}),\, \boldsymbol{\Upsilon}_L,\, \boldsymbol{\Upsilon}_{L}^{\scriptscriptstyle TP} \;\big|\; col_i(\boldsymbol{q}_{\scriptscriptstyle TP})\Big).
\end{aligned}
$}
\end{equation}
This assumption holds when APs are sufficiently spaced apart, ensuring that shadowing and multipath effects are uncorrelated across APs, as is typically the case in practical deployments~\cite{cell_free_mimo_book}.
We can further apply Bayes’ rule to each term in the product in (\ref{eqn19}) and reduce it to 
\begin{equation}
\label{eqn20}
\resizebox{\columnwidth}{!}{$
\begin{aligned}
&p\Big(col_i(\boldsymbol{q}_{\scriptscriptstyle TP}) \;\big|\; col_i(\boldsymbol{Q}_{\scriptscriptstyle RP}),\, \boldsymbol{\Upsilon}_1,\, \boldsymbol{\Upsilon}_{1}^{\scriptscriptstyle TP},\, \dots,\, \boldsymbol{\Upsilon}_L,\, \boldsymbol{\Upsilon}_{L}^{\scriptscriptstyle TP} \Big) =  \\
&\frac{ \prod_{\ell=1}^{L} p\Big(col_i(\boldsymbol{q}_{\scriptscriptstyle TP}) \;\big|\; col_i(\boldsymbol{Q}_{\scriptscriptstyle RP}),\, \boldsymbol{\Upsilon}_\ell,\, \boldsymbol{\Upsilon}_{\ell}^{\scriptscriptstyle TP} \Big) \;\prod_{\ell=1}^{L} p\Big(col_i(\boldsymbol{Q}_{\scriptscriptstyle RP}),\, \boldsymbol{\Upsilon}_\ell,\, \boldsymbol{\Upsilon}_{\ell}^{\scriptscriptstyle TP}\Big) }{ p\Big(col_i(\boldsymbol{q}_{\scriptscriptstyle TP})\Big)^{L-1} \; p\Big(col_i(\boldsymbol{Q}_{\scriptscriptstyle RP}),\, \boldsymbol{\Upsilon}_1,\, \boldsymbol{\Upsilon}_{1}^{\scriptscriptstyle TP},\, \dots,\, \boldsymbol{\Upsilon}_L,\, \boldsymbol{\Upsilon}_{L}^{\scriptscriptstyle TP} \Big) }.
\end{aligned}
$}
\tag{20}
\end{equation}
The UE has access to the conditional distributions $p\left(col_i(\boldsymbol{q}_{\scriptscriptstyle TP}) \;\big|\; col_i(\boldsymbol{Q}_{\scriptscriptstyle RP}),\, \boldsymbol{\Upsilon}_\ell,\, \boldsymbol{\Upsilon}_{\ell}^{\scriptscriptstyle TP} \right)$, each of which is shown to be Gaussian with mean $\mu_{\scriptscriptstyle i\ell}$ and variance $v_{\scriptscriptstyle i\ell}$ in \Cref{GPR_based_loc}. We further assume a non-informative prior for $p\left(col_i(\boldsymbol{q}_{\scriptscriptstyle TP})\right)$, modeled as a Gaussian distribution with a large variance. Under these assumptions, the joint posterior position estimate $
p\left(col_i(\boldsymbol{q}_{\scriptscriptstyle TP}) \;\big|\; col_i(\boldsymbol{Q}_{\scriptscriptstyle RP}),\, \boldsymbol{\Upsilon}_1,\, \boldsymbol{\Upsilon}_{1}^{\scriptscriptstyle TP},\, \dots,\, \boldsymbol{\Upsilon}_L,\, \boldsymbol{\Upsilon}_{L}^{\scriptscriptstyle TP} \right)$
is proportional to the product of $L$ Gaussian probability density functions, each corresponding to 
$p(col_i(\boldsymbol{q}_{\scriptscriptstyle TP}) \mid col_i(\boldsymbol{Q}_{\scriptscriptstyle RP}),\, \boldsymbol{\Upsilon}_\ell,\, \boldsymbol{\Upsilon}_{\ell}^{\scriptscriptstyle TP})$.
This results in a Gaussian distribution with combined variance and mean,  given respectively by~\cite{gaussian_products}:
\begin{equation}
\label{eqn21}
\tag{21}
v_i = \left( \sum_{\ell=1}^L \frac{1}{v_{i\ell}} \right)^{-1}, \quad
\mu_i = v_i \left( \sum_{\ell=1}^L \frac{\mu_{\scriptscriptstyle i\ell}}{v_{i\ell}} \right).
\end{equation}
While not influencing the final Gaussian-distributed position estimate in the distributed-Bayesian algorithm itself, the term $
\frac{\prod_{\ell=1}^{L} p\big(col_i(\boldsymbol{Q}_{\scriptscriptstyle RP}),\, \boldsymbol{\Upsilon}_\ell,\, \boldsymbol{\Upsilon}_{\ell}^{\scriptscriptstyle TP}\big)}{p\big(col_i(\boldsymbol{q}_{\scriptscriptstyle TP})\big)^{L-1} \, p\big(col_i(\boldsymbol{Q}_{\scriptscriptstyle RP}),\, \boldsymbol{\Upsilon}_1,\, \boldsymbol{\Upsilon}_{1}^{\scriptscriptstyle TP},\, \dots,\, \boldsymbol{\Upsilon}_L,\, \boldsymbol{\Upsilon}_{L}^{\scriptscriptstyle TP} \big)}
$ in (\ref{eqn20}), serves as a normalizing constant, ensuring that the total probability integrates to one.

The final position estimate $\overline{\boldsymbol{q}}_{\scriptscriptstyle DB} = \overline{\boldsymbol{q}}_{est}$ = $(\overline{x}_{\scriptscriptstyle DB}, \overline{y}_{\scriptscriptstyle DB})$ =  $(\mu_1,\mu_2) \in \mathbb{R}^{1 \times 2}$ with associated variance $\boldsymbol{v}_{\scriptscriptstyle DB}$ = $(v^{\scriptscriptstyle DB}_1,v^{\scriptscriptstyle DB}_2)$ = $({v}_{1},{v}_{2}) \in \mathbb{R}^{1 \times 2}$ achieves the minimum variance among the proposed algorithms (Appendix \ref{appendixa}). 

\subsubsection{Distributed-Z-Score Algorithm} The distributed-Bayesian approach typically achieves low variance; however, it weighs the means of all APs to compute the final estimate, including those with poor RSS and AOA estimates due to high shadowing noise. Consequently, these inaccurate estimates, acting as outliers, can degrade the overall localization accuracy. By mitigating the influence of these outliers, the distributed-z-score algorithm can potentially achieve better localization accuracy. It introduces a statistical filtering step based on the z-score, which quantifies the number of standard deviations a data point is from the mean. For each coordinate, the z-score of the 
$L$ AP's MMSE estimates is calculated. Only the values with z-scores within a specified threshold are retained. The UE subsequently applies the Bayesian principles to these filtered values to obtain the final system estimate of the coordinate. This process is detailed in Algorithm~\ref{algo:zscore_coordinates}.
\begin{algorithm}[h]
  \caption{Distributed-Z-Score Algorithm}
  \label{algo:zscore_coordinates}
  
  \textbf{Input:} List of coordinate estimates \( \overline{\boldsymbol{q}}^{{est}}_\ell = (\overline{x}_\ell^{{est}}, \overline{y}_\ell^{{est}}) \), variances \( {v}_{\scriptscriptstyle i\ell} \), for \( \ell = 1, 2, \ldots, L \) \\
  \textbf{Output:} Filtered coordinates \( \overline{\boldsymbol{q}}_{\scriptscriptstyle DZ} = (\overline{x}_{\scriptscriptstyle DZ}, \overline{y}_{\scriptscriptstyle DZ}) \), variances \( \boldsymbol{v}_{\scriptscriptstyle DZ} = (v^{\scriptscriptstyle DZ}_1, v^{\scriptscriptstyle DZ}_2) \)
  
  \begin{algorithmic}[1]
    \State Compute mean \( \mu_x \) and standard deviation \( \sigma_x \) of \( \{\overline{x}_\ell^{{est}}\} \)
    \State Compute mean \( \mu_y \) and standard deviation \( \sigma_y \) of \( \{\overline{y}_\ell^{{est}}\} \)
    
    \For{\( \ell = 1 \) to \( L \)}
      \State Compute z-score for \( x \)-coordinate: \( Z^x_\ell \gets \frac{\overline{x}_\ell^{{est}} - \mu_x}{\sigma_x} \)
      \State Compute z-score for \( y \)-coordinate: \( Z^y_\ell \gets \frac{\overline{y}_\ell^{{est}} - \mu_y}{\sigma_y} \)
    \EndFor
     \State Set z-score threshold $\mathcal{T}_z$
    \State Define filtered index sets:
    \[
    \mathcal{L}_x = \{\ell \mid |Z^x_\ell| < \mathcal{T}_z\}, \quad
    \mathcal{L}_y = \{\ell \mid |Z^y_\ell| < \mathcal{T}_z\}
    \]
    
    \State Compute combined variances:
    \[
    v_1^{\scriptscriptstyle DZ} \gets \left( \sum_{\ell \in \mathcal{L}_x} \frac{1}{{v}_{\scriptscriptstyle 1\ell}} \right)^{-1}, \quad
    v_2^{\scriptscriptstyle DZ} \gets \left( \sum_{\ell \in \mathcal{L}_y} \frac{1}{{v}_{\scriptscriptstyle 2\ell}} \right)^{-1}
    \]
    
    \State Compute weighted means:
    \[
    \overline{x}_{\scriptscriptstyle DZ} \gets v_1^{\scriptscriptstyle DZ} \sum_{\ell \in \mathcal{L}_x} \frac{\overline{x}_\ell^{{est}}}{{v}_{\scriptscriptstyle 1\ell}}, \quad
    \overline{y}_{\scriptscriptstyle DZ} \gets v_2^{\scriptscriptstyle DZ} \sum_{\ell \in \mathcal{L}_y} \frac{\overline{y}_\ell^{{est}}}{{v}_{\scriptscriptstyle 2\ell}}
    \]
    
    \State Return \( \overline{\boldsymbol{q}}_{\scriptscriptstyle DZ} \gets (\overline{x}_{\scriptscriptstyle DZ}, \overline{y}_{\scriptscriptstyle DZ}) \), \( \boldsymbol{v}_{\scriptscriptstyle DZ} \gets (v^{\scriptscriptstyle DZ}_1, v^{\scriptscriptstyle DZ}_2) \)
  \end{algorithmic}
\end{algorithm}

The z-score threshold $\mathcal{T}_z$ is a hyperparameter selected according to the desired estimation quality. A larger $\mathcal{T}_z$ results in fewer coordinate estimates being filtered out, potentially including outliers. However, with more coordinate estimates contributing to the final estimate, the variance of the final estimate decreases, leading to reduced uncertainty. Conversely, a lower $\mathcal{T}_z$ filters out more AP estimates, potentially excluding valuable ones, and increases variance due to fewer APs contributing to the final weighted mean estimate. Therefore, by careful selection of $\mathcal{T}_z$ we can eliminate APs with outlier estimates, resulting in $\overline{\boldsymbol{p}}_{\scriptscriptstyle DZ}$ being more accurate than its bayesian counterpart $\overline{\boldsymbol{p}}_{\scriptscriptstyle DB}$. Consequently, $v^{\scriptscriptstyle DZ}_i$ would also be comparable to $v^{\scriptscriptstyle DB}_i$, both of which are significantly lower than $v^{\scriptscriptstyle DD}_i$ (see Appendix \ref{appendixa}).
\section[Complexity Analysis and Performance Metrics]{Complexity Analysis and Performance Metrics}
\label{complexity_and_performance}
In this section, we analyze the computational complexity of the proposed distributed algorithms and compare them to the centralized GPR method (centralized-hybrid), which leverages RSS and AOA from all APs for localization at the CPU, as introduced in~\cite{my_wcl_letter}. Additionally, we discuss the performance metrics used in Section \ref{sim_results} for evaluating the algorithms.

\vspace{\vertspacebeforesubsections}
\subsection{Complexity Analysis of the Offline Phase}
The offline phase computations include calculating RSS vectors for each AP, estimating hyperparameters and error variances, and inverting the covariance matrices of the training vectors as described in (\ref{eqn17}).  The AOA vectors do not contribute to the computational load on the APs in the offline phase, as they are precomputed using geometrical methods and directly incorporated into the AP's memory.

If the received signal at AP $\ell$ is averaged over $\mathzapf{S_{\scriptscriptstyle Y}}$ samples, the complexity of RSS calculation per AP, per RP is $\mathbf{\mathcal{O}}(\mathzapf{S_{\scriptscriptstyle Y}}Nz)$, as shown in \eqref{eqn5}. Therefore, the total complexity across all $K$ RPs is $\mathbf{\mathcal{O}}(\mathzapf{S_{\scriptscriptstyle Y}}KNz)$ for each AP in the distributed framework, whereas in the centralized-hybrid case, the CPU must compute the RSS for all APs, resulting in a complexity of $\mathbf{\mathcal{O}}(\mathzapf{S_{\scriptscriptstyle Y}}KLNz)$. 

Using basic gradient ascent, the optimization problem described in (\ref{eqn13}) can be solved as 
\begin{equation} 
\tag{22}
\label{eqn22}
{\boldsymbol{\Gamma}}_{\scriptstyle i\ell}^{n+1} = {\boldsymbol{\Gamma}}_{\scriptstyle i\ell}^{n} + \eta \nabla_{\scriptscriptstyle  \boldsymbol{\Gamma}_{\scriptscriptstyle i\ell}} \log\big(p\left(\operatorname{col}_i(\boldsymbol{Q}_{\scriptscriptstyle RP}) \mid \boldsymbol{\Upsilon}_\ell, \boldsymbol{\Gamma}^n_{\scriptstyle i\ell} \right)\big),
\end{equation}
where ${\boldsymbol{\Gamma}}_{\scriptscriptstyle i\ell}^{n}$ denotes the estimate of $\boldsymbol{\Gamma}_{\scriptscriptstyle i\ell}$ at the $n^{th}$ iteration of the gradient ascent. The learning rate $\eta$ is a hyperparameter that controls the step size during the optimization. The gradient of the log-likelihood is computed as~\cite{gpr_book_mit},
\begin{align}
&\nabla_{\scriptstyle \boldsymbol{\Gamma}_{\scriptstyle i\ell}} 
\log\left(p\left(\text{col}_i(\boldsymbol{Q}_{\scriptscriptstyle RP}) \mid \boldsymbol{\Upsilon}_{\ell}, \boldsymbol{\Gamma}_{\scriptstyle i\ell}\right)\right)
\nonumber \\ 
= &
\frac{1}{2} \, \text{tr} \left( \left( \boldsymbol{\Lambda}_{i\ell} \boldsymbol{\Lambda}_{i\ell}^\top 
- \widetilde{\bm{\mathcal{K}}_{i\ell}}^{-1} \right) 
\nabla_{\scriptscriptstyle \boldsymbol{\Gamma}_{\scriptstyle i\ell}} \widetilde{\bm{\mathcal{K}}_{i\ell}} \right)
\tag{23}
\label{eqn23}
\end{align}
where $\widetilde{\bm{\mathcal{K}}}_{\scriptstyle i\ell} = [\bm{\mathcal{K}}_{\scriptstyle i\ell}(\boldsymbol{\Upsilon}_{\ell},\boldsymbol{\Upsilon}_{\ell}) + \sigma_{\epsilon_{\scalebox{0.5}{$i\ell$}}}^2\textbf{I}_K]$ is the regularized covariance matrix and $\boldsymbol{\Lambda}_{\scriptstyle i\ell} = \widetilde{\bm{\mathcal{K}}_{\scriptstyle i\ell}}^{-1} [col_i(\boldsymbol{Q}_{\scriptscriptstyle RP})]$.
The optimization process begins with an initial estimate of the learnable parameters and iterates until the change in the log-likelihood (or the hyperparameters) falls below a predefined threshold. 

The computation of each covariance matrix $\bm{\mathcal{K}}_{\scriptstyle i\ell}(\boldsymbol{\Upsilon}_{\ell},\boldsymbol{\Upsilon}_{\ell})$ and the associated matrix multiplications in (\ref{eqn23}) have a complexity of  $\mathbf{\mathcal{O}}(K^2)$, while inverting each $\widetilde{\bm{\mathcal{K}}}_{i\ell}$ requires $\mathbf{\mathcal{O}}(K^3)$, resulting in an overall complexity of $\mathbf{\mathcal{O}}(K^3+K^2)$ per AP, per iteration. For the centralized-hybrid approach, the CPU computes the covariance matrix with a complexity of $\mathbf{\mathcal{O}}(LK^2)$ as each matrix entry depends on vector norms involving RSS and AOA values from all APs. Including the covariance matrix inversion step, the total complexity per iteration for the centralized-hybrid case becomes $\mathbf{\mathcal{O}}(K^3 + LK^2)$.

Since the squared exponential kernel is Lipschitz continuous~\cite{gpr_lipschitz}, the gradient ascent optimization in (\ref{eqn22}) follows an iteration complexity bound of $\mathbf{\mathcal{O}}\left(\omega_{\text{distr}}/{\tau}\right)$, where $\omega_{\text{distr}}$ is the Lipschitz constant of the log-likelihood function in the distributed setting, and $\tau$ is the predefined convergence threshold on the norm of the gradient. Similarly, for the centralized-hybrid approach, the iteration complexity is $\mathbf{\mathcal{O}}\left(\omega_{\text{centr}}/{\tau}\right)$, where $\omega_{\text{centr}}$ is the corresponding Lipschitz constant. Consequently, the total offline phase complexity for the distributed and centralized-hybrid approaches is $\mathbf{\mathcal{O}}\left(\mathzapf{S_{\scriptscriptstyle Y}}KNz + {\omega_{\text{distr}}}(K^3 + K^2)/{\tau} \right)$ and $\mathbf{\mathcal{O}}\left(\mathzapf{S_{\scriptscriptstyle Y}}KLNz + {\omega_{\text{centr}}}(K^3 + LK^2)/{\tau} \right)$ respectively.

It is well known that the Lipschitz constant depends on the dimensionality of the input space \cite{lipschitz_dimension}. In the centralized-hybrid case, the input space dimensionality is $L$ times larger than in the distributed case, as the input vector at the CPU is a concatenation of RSS and AOA vectors from all APs. Consequently, the centralized-hybrid approach incurs higher computational complexity per unit compared to the distributed approach in the offline phase.

\vspace{\vertspacebeforesubsections}
\subsection{Complexity Analysis of the Online Phase}
In the online phase, the computational complexity primarily involves calculating the test point vector including the RSS values and the AOA values, forming the covariance vector of the training and test data, and performing the matrix operations specified in (\ref{eqn17}). Since the RSS is computed using the same method as in the offline phase, its complexity per AP in the distributed approach is $\mathbf{\mathcal{O}}(\mathzapf{S_{\scriptscriptstyle Y}}Nz)$, whereas in the centralized-hybrid case, for the CPU it is $\mathbf{\mathcal{O}}(\mathzapf{S_{\scriptscriptstyle Y}}LNz)$.

The AOA computation using MUSIC involves estimating the covariance matrix $\textbf{R}_{\scriptscriptstyle TP,\ell}$ with complexity $\mathbf{\mathcal{O}}(\mathzapf{S_{\scriptscriptstyle Y}}Nz)$, computing its eigenvectors with $\mathbf{\mathcal{O}}(N^3)$ and evaluating the MUSIC pseudospectrum $\mathrsfso{P}_{\scriptscriptstyle MUSIC}(\theta)$ over $\mathrsfso{N}_{\scriptscriptstyle srch}$ search points to find the peak, requiring $\mathbf{\mathcal{O}}(\mathrsfso{N}_{\scriptscriptstyle srch}N^2)$ for each AP~\cite{MUSIC_complexity}. Thus, the total complexity per AP is $\mathbf{\mathcal{O}}(N^3 + \mathrsfso{N}_{\scriptscriptstyle srch}N^2 + \mathzapf{S_{\scriptscriptstyle Y}}Nz)$, while for the centralized-hybrid approach it is $\mathbf{\mathcal{O}}(L(N^3 + \mathrsfso{N}_{\scriptscriptstyle srch}N^2 + \mathzapf{S_{\scriptscriptstyle Y}}Nz))$ at the CPU, as it must calculate the AOAs for all APs.

Computation of the covariance vectors $\bm{\mathcal{K}}_{\scriptstyle i\ell}(\boldsymbol{\Upsilon}_{\ell},\boldsymbol{\Upsilon}_{\ell}^{\scriptscriptstyle TP})$ requires  $\mathbf{\mathcal{O}}(K)$ operations, while the matrix products in (\ref{eqn17}) add an additional $\mathbf{\mathcal{O}}(K^2)$, resulting in a total complexity of $\mathbf{\mathcal{O}}(K^2+K)$ per AP in the distributed setting for model evaluation. Following the same reasoning as in the offline phase, for the centralized-hybrid approach, the total complexity for model evaluation is $\mathbf{\mathcal{O}}(K^2+LK)$ for the CPU. 

Unlike in the centralized-hybrid algorithm, in distributed localization, in addition to each AP estimating the position using the learned GPR model, the UE must also aggregate these estimates to compute the final position using the described algorithmic variant. The operations for each variant (median, mean, Bayesian, and z-score) require $\mathbf{\mathcal{O}}(L)$ operations at the UE. Since $L \ll K$  in practical localization systems, the online phase complexity at each AP is effectively $\mathcal{O}(K^2)$, which is quadratic in the database size and well within the processing capabilities of modern network infrastructure~\cite{my_wcl_letter}. For the UE, the aggregation step operates in effectively constant time. Therefore, even in the online phase, the total complexity per network-side computing unit is lower in the proposed distributed framework compared to the centralized framework.

\vspace{\vertspacebeforesubsections}
\subsection{The Performance Metrics}
To evaluate the localization accuracy we utilize the average Euclidean distance relative to the true coordinates of a test point where a UE is positioned for estimation. Formally, the localization error $q_{err}$ is calculated as:
\begin{equation}
\tag{24}
\label{eqn24}
\resizebox{0.89\columnwidth}{!}{$
\begin{split}
q_{err} &= \|\boldsymbol{q}_{\scriptscriptstyle TP} - \overline{\boldsymbol{q}}_{est}\| = \sqrt{(x_{\scriptscriptstyle TP}-\overline{x}_{\scriptscriptstyle est})^2+(y_{\scriptscriptstyle TP}-\overline{y}_{\scriptscriptstyle est})^2}.
\end{split}
$}
\end{equation}
To assess the overall uncertainty in the estimated 2D position we can calculate the area of the
95\% error ellipse which represents the region within which there is a 95\% probability that the true position lies. The area of this ellipse provides a measure of the overall positional uncertainty, combining the individual uncertainties of both the coordinates into a single metric. Mathematically, the area of the 95\% error ellipse is given by
\begin{equation}
\tag{25} 
\label{eqn25}
\mathcal{A}_{\scriptscriptstyle 95} = 5.991\pi\sqrt{v_1v_2}
\end{equation}
Derivation of the ellipse is provided in Appendix \ref{appendixb}. A smaller ellipse area indicates lower average variance across both coordinates, thereby providing greater confidence in the predicted position.

\vspace{\vertspacebeforesections}
\section{Simulation Results}
\label{sim_results}
In this section, we present a comprehensive evaluation of the proposed algorithms based on the performance metrics outlined in Section \ref{complexity_and_performance}. We compare the performance of the proposed distributed algorithms both among themselves and against established methods, including the centralized approach in \cite{my_wcl_letter}, the centralized KNN and LR techniques from \cite{fog_massive_mimo_ml} and the centralized 7-layer fully connected neural network (FCNN) based ML-MUSIC localization approach described in \cite{distributed_music_positioning}. To further benchmark GPR model, we adapt the KNN and LR models to the proposed distributed architecture. We first present the cell-free system parameters employed in the simulations and subsequently discuss the regression results. All simulation code and results are provided in the repository linked in~\cite{github_code_link}.

\vspace{\vertspacebeforesubsections}
\subsection{Cell-Free System Parameters}
The APs are deployed randomly in an urban environment of network area 200m$\times$200m. RPs are systematically arranged in a square pattern, forming a grid that spans the entire simulation area, as shown in Fig. \ref{system_model_fig}. Each large-scale fading coefficient is calculated using (\ref{eqn5}) with $p^{0}_{\ell}=-28.8\ dB$ at $d_{\ell}^{0} = 1m$ and $\gamma=3.53$. These parameters are derived from the 3rd generation partnership project 38.901 urban micro street canyon NLOS path loss model~\cite{3gpp_38_901}. The shadowing terms from an AP to distinct location points in the network area are correlated as~\cite{cell_free_mimo_book}
\begin{equation}
\tag{26}
\label{eqn26}
\mathbb{E}\{\nu_{m\ell}\nu_{ij}\} = 
\begin{cases} 
\sigma_{\scriptscriptstyle SF}^2 \cdot 2^{-\frac{d_{\scriptscriptstyle mi}}{d_{\scriptscriptstyle corr}}}, & \text{if } \ell = j, \\
    0, & \text{otherwise}.
\end{cases}
\end{equation}
Here, $\nu_{m\ell}$ is the shadowing from AP $\ell$ to location point $m$, $d_{mi}$ is the distance between locations $i$ and $m$ and $d_{corr}$ is the decorrelation distance that is characteristic of the environment. The location point indices $m$ correspond to the RPs and the test point during a localization exercise.

The scatterers surrounding the UE are modeled as being distributed in accordance with the disk scattering model. For this model, the covariance matrix $\textbf{R}^{ds}_{\scriptscriptstyle TP,\ell}\in \mathbb{R}^{N \times N}$ of the received signal at AP $\ell$, for $z=1$ is given by \cite{disk_scatter_model}
\begin{equation}
\tag{27}
\label{eqn27}
\textbf{R}^{ds}_{\scriptscriptstyle TP,\ell} = \rho\beta_{\scriptscriptstyle TP,\ell}\textbf{G}(\zeta)\odot \textbf{a}(\varphi_{\scriptscriptstyle TP,\ell}^{\circ})\textbf{a}^H(\varphi_{\scriptscriptstyle TP,\ell}^{\circ}) + \sigma_n^2\textbf{I}_N.
\end{equation}
Here $\textbf{G}(\zeta) \in \mathbb{R}^{N \times N}$ is the matrix of scaling factors with elements $\begin{bmatrix}
\textbf{G}(\zeta)
\end{bmatrix}_{mn} = [J_0((m-n)\zeta) + J_2((m-n)\zeta)]$,  $\zeta = \frac{2\pi d}{\lambda}\Delta\sin(\varphi_{\scriptscriptstyle TP,\ell}^{\circ})$, $\Delta$ is the single side angular spread of the multipath components impinging the AP antenna array according to the disk scattering model and $J_k$ is the Bessel function of the first kind and order $k$. Eq. (\ref{eqn27}) is valid for small angular spreads such that $\sin(\Delta) \approx \Delta$.

Given that AOA is intended to be measured through geometrical methods in the offline stage, we introduce a nominal error of $\mathcal{N}(0,4)$ degree to the true azimuth AOA of every AP. This serves to model the inherent measurement errors encountered in practical applications. Also, we disregard the 180-degree ambiguity inherent to ULAs, assuming it can be resolved, for example, by employing spatial array processing techniques \cite{180_degree_ambiguity}. Table \ref{tab:table1} provides additional simulation parameters~\cite{cell_free_mimo_book}.

\begin{table}[!t]
\caption{Simulation Parameters\label{tab:table1}}
\centering
\begin{tabular}{|c|c|}
\hline
\textbf{Parameter} & \textbf{Value}\\
\hline
Carrier frequency & 2 GHz\\
\hline
Signal bandwidth & 10 MHz\\
\hline
Number of APs ($L$) & 25 \\
\hline
Antenna array spacing ($d$) & $0.5\lambda$ \\
\hline
Height of an AP & 10 m \\
\hline
Height of the UE & 1.5 m \\
\hline
UE transmit power ($\rho$) & 100 mW \\
\hline
Noise power ($\sigma_n^2$) & -96 dBm \\
\hline
Receiver noise figure & 8 dB \\
\hline
Decorrelation distance ($d_{\scriptscriptstyle corr}$) & 13 m \\
\hline
Number of distinct setups, each with randomly placed APs & 100\\
\hline
Number of randomly placed test points per setup & 1000\\
\hline
Received signal samples used for estimating RSS and  $\textbf{R}^{ds}_{\scriptscriptstyle TP,\ell}$ & 200 \\
\hline

Single side angle spread ($\Delta$)& 10\textdegree \\
\hline
\end{tabular}
\vspace{-4mm}
\end{table}

\vspace{\vertspacebeforesubsections}
\subsection{Simulation Results and Discussions}
\subsubsection{Localization Accuracy for Varying AP Antenna Counts}
\begin{figure}[!t]
\centering
\subfloat[$K=64$]{%
    \includegraphics[width=2.5in]{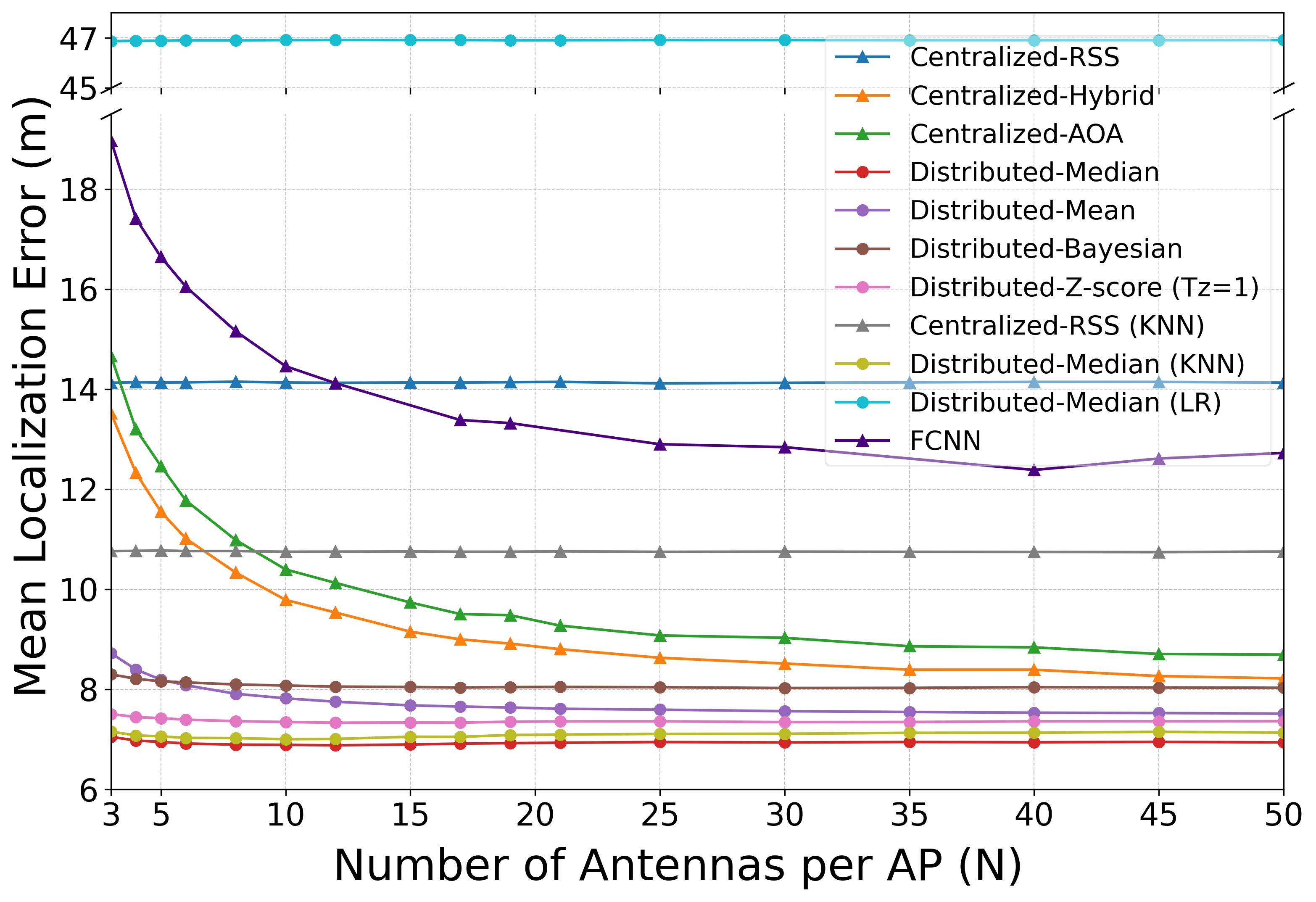}%
    \label{fig:n_vs_err_a}%
}
\vspace{-1em}
\subfloat[$K=225$]{%
    \includegraphics[width=2.4in]{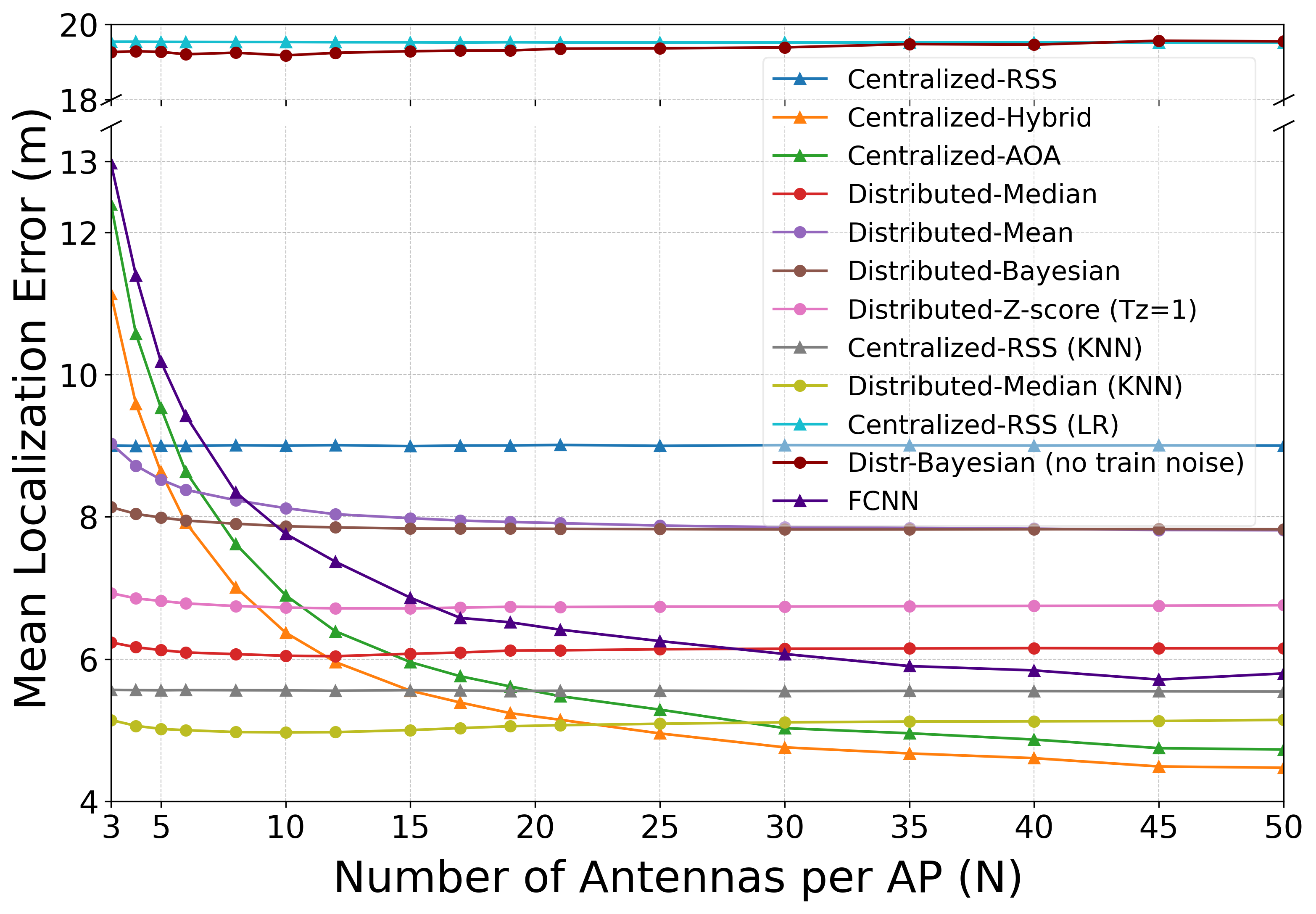}%
    \label{fig:n_vs_err_b}%
}

\caption{Localization accuracy for varying number of AP antennas ($N$) for $L=25,\ \sigma_{\scriptscriptstyle SF}=8\,\mathrm{dB}$, $\mathcal{T}_z=1$.}
\label{n_vs_est_err}
\end{figure}

Fig. \ref{n_vs_est_err} presents the average localization error across 100 different setups, with 1000 test points per setup, as a function of various values of $N$ for the proposed regression methods. The results are presented for two values of $K$: 64 and 225, representing different RP densities within the simulation area. The centralized GPR approach, along with the KNN, LR, and FCNN methods, serve as baseline comparisons. The FCNN takes the $L$ dimensional AOA vector from $L$ access points as input, passes it through five hidden layers with sizes 128, 64, 32, 32 and 16 using hyperbolic tangent activations, and outputs the estimated $x$ and $y$ coordinates via a linear output layer. Training is performed on the aggregated AOA fingerprint database at the CPU using the scaled conjugate gradient algorithm for 500 epochs. In addition, the KNN and LR algorithms are applied to the distributed-median framework introduced in \Cref{UE_aggregation}. Here, the GPR is replaced with KNN and LR, where each AP performs local regression with KNN and LR respectively, using its own hybrid inputs of RSS and AOA. The UE estimates the final position by taking the median of the AP predictions. 

For $K=64$ the distributed algorithms achieve the highest localization accuracy, whereas for 
$K=225$, the centralized-hybrid and AOA GPRs deliver the best performance. At lower values of $K$, the degraded performance of centralized GPR can be attributed to the curse of dimensionality~\cite{curse_and_reg}.
Given input vector lengths of $2L=50$ for the centralized-hybrid GPR and $L=25$ for the centralized-AOA and RSS GPRs, the available $K=64$ training points are inadequate for effective learning in high-dimensional spaces. In contrast, with $K=225$, the centralized models benefit from a larger training dataset, allowing them to leverage information from all APs simultaneously. This facilitates the learning of a unified model that captures richer spatial and signal characteristics, effectively exploiting inter-AP correlations and thereby outperforming distributed algorithms, where models are trained independently at each AP.

Further, we can observe a general decreasing trend in localization error for algorithms utilizing AOA inputs, since the MUSIC algorithm's accuracy improves with an increasing number of antennas~\cite{MUSIC_with_ant}. This trend is more pronounced in centralized approaches, as the test vector in these cases is higher-dimensional, incorporating AOA from all APs, which improves with increasing antenna count. In contrast, the distributed algorithms use a two-dimensional test vector consisting of RSS and AOA of individual AP, where only AOA benefits from additional antennas. This limited improvement in feature richness is insufficient to significantly improve the localization accuracy with increasing $N$.

Among the distributed schemes, the distributed-median algorithm yields the highest localization accuracy for both values of $K$, outperforming the distributed-mean, Bayesian, and z-score methods.  However, as discussed earlier in Section \ref{UE_aggregation} and as we will explore further in the uncertainty analysis, the distributed-mean and Bayesian algorithms exhibit significantly lower variance. The distributed-z-score algorithm achieves a balance by delivering accuracy close to the distributed-median while maintaining low variance comparable to the Bayesian and mean methods. Interestingly, KNN performs comparably well for both values of \( K \), despite its simplicity. However, its performance has reached a saturation, as it relies on a basic weighted average of neighboring points. Given the straightforward nature of the algorithm, further improvements are limited to adjusting the number of neighbors $k$, which has already been shown to be optimal at $k=4$ in~\cite{fog_massive_mimo_ml}. Consequently, further gains would depend on external factors such as reduced shadowing noise or increased RP density, which are outside the user’s control. In contrast, the GPR method retains potential for enhancement through kernel design, hyperparameter tuning, and improved AOA estimation.

\begin{figure}
\centering
\captionsetup{justification=centering}
\includegraphics[width=2.4 in]{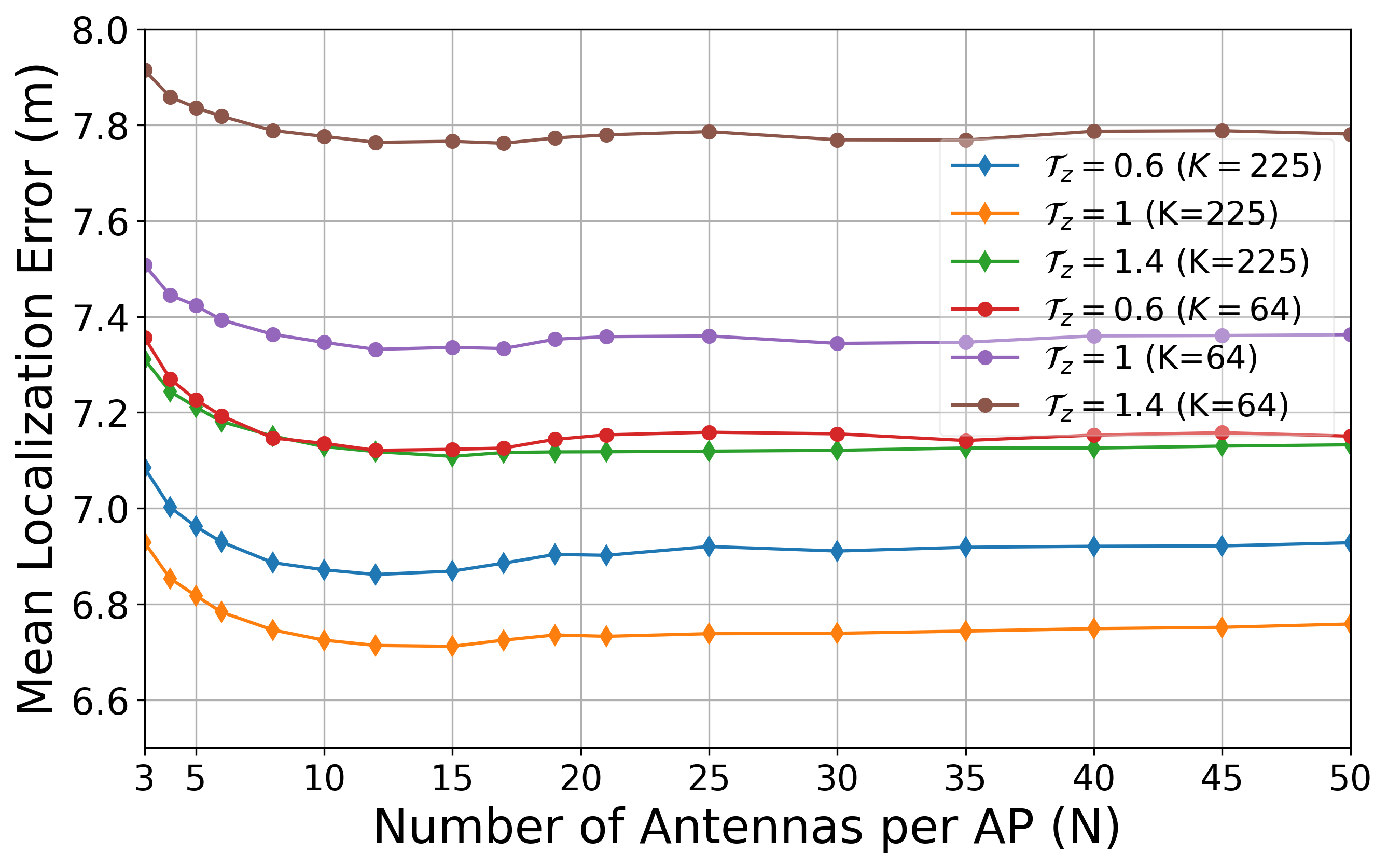}
\caption{Localization accuracy for different distributed-z-score thresholds $\mathcal{T}_z$, for $L=25,\ \sigma_{\scriptscriptstyle SF}=8dB$.}
\label{n_vs_est_err_Tz}
\end{figure}

Fig. \ref{n_vs_est_err_Tz} shows the localization performance of the distributed-z-score algorithm as a function of $N$ for different threshold values $\mathcal{T}_z$. While the general decreasing trend in localization error persists across all threshold values, the mean localization error is the lowest when $\mathcal{T}_z$ is set to 1, for the corresponding $K$. In this case, APs with mean estimates deviating by more than one standard deviation are excluded, effectively filtering out outliers. For $\mathcal{T}_z>1$, some outlier APs are retained, slightly increasing the error, while $\mathcal{T}_z<1$ removes reliable APs, leading to higher errors. Furthermore, unlike ~\cite{ml_rss_surya,analytical_surya}, where the GPR model is trained with noiseless inputs, our results show that retaining noise during training improves localization accuracy. This is due to regularization, which mitigates model overfitting and enhances generalization~\cite{curse_and_reg}.

\subsubsection{95\% Error Ellipse Analysis of Position Estimates}
\begin{figure}[!t]
\centering
\subfloat[$K=64$]{%
    \includegraphics[width=2.5in]{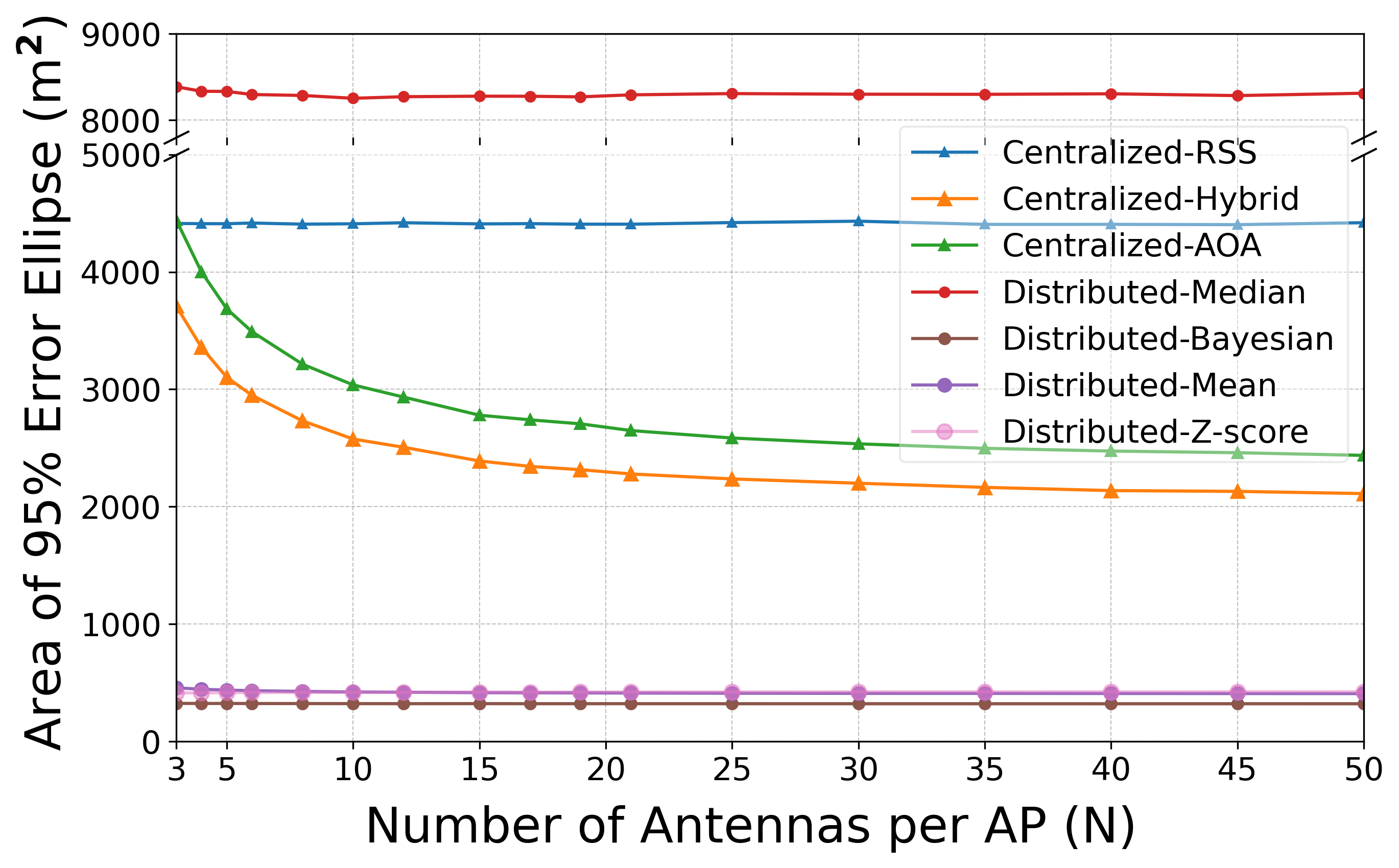}%
    \label{fig:n_vs_ellipse_a}%
}
\vspace{-1em}
\subfloat[$K=225$]{%
    \includegraphics[width=2.5in]{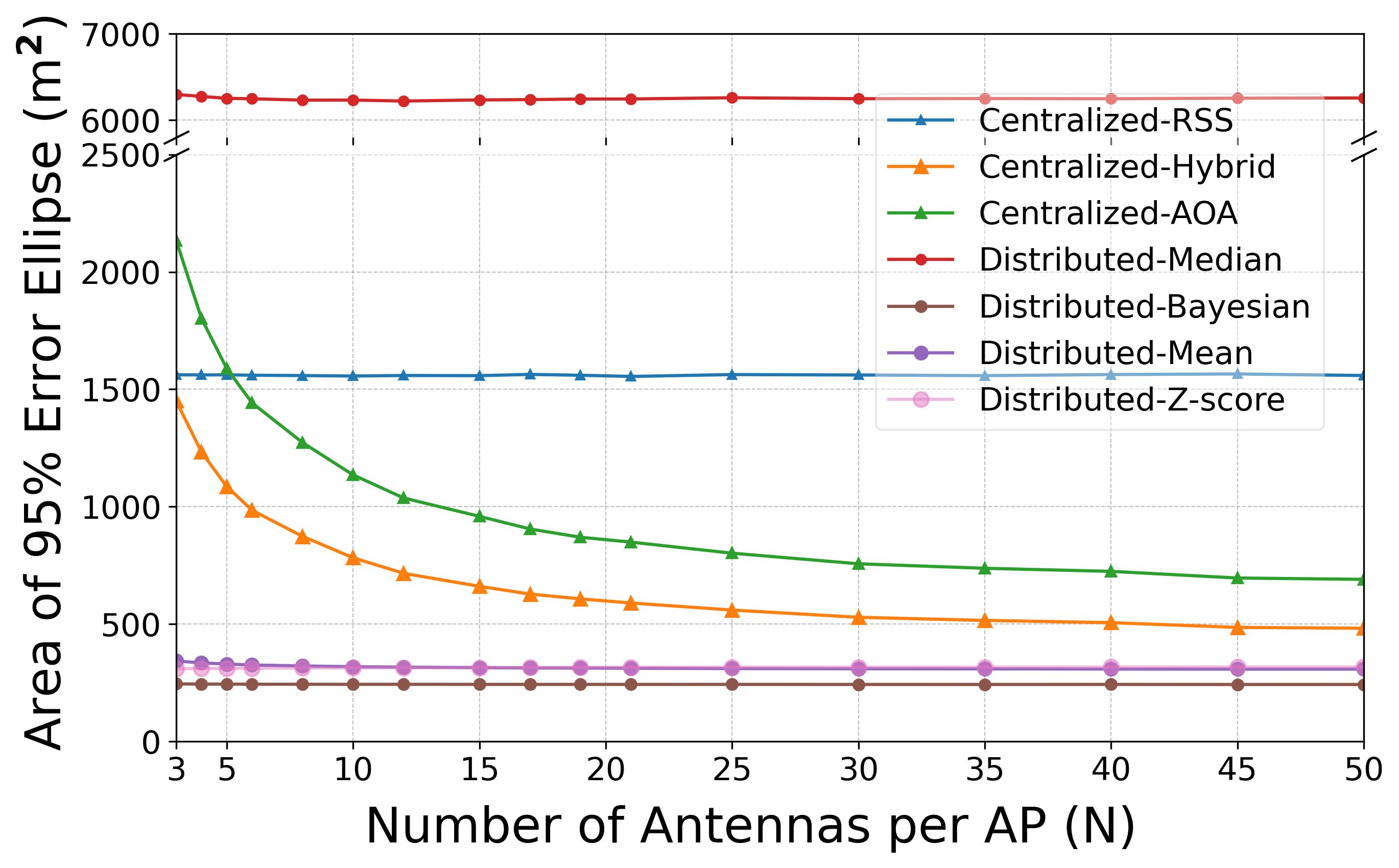}%
    \label{fig:n_vs_ellipse_b}%
}

\caption{Area of 95\% error ellipse for varying number of antennas ($N$) per AP for $L=25,\ \sigma_{SF}=8dB$, $\mathcal{T}_z=1$.}
\label{n_vs_ellipse}
\end{figure}

The distributed-median algorithm exhibits the highest uncertainty despite achieving the lowest localization error for both values of $K$. Fig. \ref{n_vs_ellipse} shows that the average ellipse area across all antennas is $6248\,\text{m}^2$ when $K=225$. This corresponds to an unreasonably high standard deviation of 18.2$\,\text{m}$ per coordinate. The high uncertainty stems from its reliance on a single AP's estimate, significantly increasing variance in the final prediction. In contrast, the distributed mean, Bayesian, and z-score algorithms aggregate estimates from multiple APs, reducing uncertainty to average ellipse areas of 317$\,\text{m}^2$, 243$\,\text{m}^2$ and 314$\,\text{m}^2$, respectively. These distributed algorithms, with an average standard deviations of 4.1$\,\text{m}$, 3.6$\,\text{m}$ and 4.1$\,\text{m}$ respectively, outperforms even the centralized variants despite lacking centralized configuration benefits, highlightling the advantage of aggregating multiple estimates. The larger ellipse area in the distributed-z-score method compared to the distributed-Bayesian method results from filtering out some AP estimates, to remove outliers for enhancing the accuracy.

\begin{figure}
\centering
\captionsetup{justification=centering}
\includegraphics[width=2.7 in]{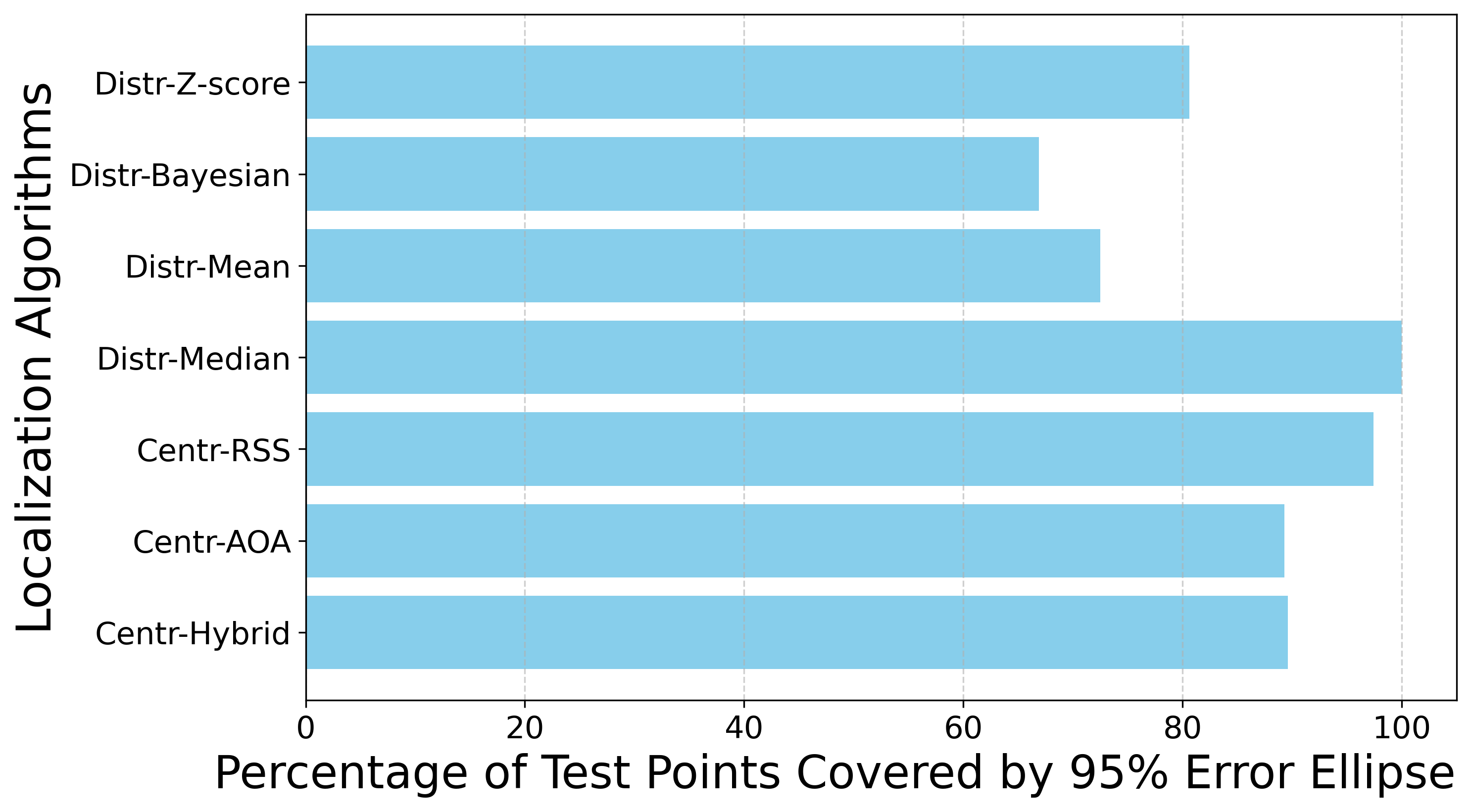}
\caption{Percentage of test points covered by 95\% error ellipse, for $L=25,\ \sigma_{SF}=8dB$, $\mathcal{T}_z=1$ and $K=225$.}
\label{percent_in_ellipse}
\end{figure}

Unlike \cite{ml_rss_surya} where only 15\% of test points fell within the $\pm 2\sigma$ error bars for conventional GPR, our observations show that, for $K=225$, 67\%, 73\%, 81\%, and 90\% of test points lie within the 95\% error ellipses for the distributed-Bayesian, distributed-mean, distributed-z-score and centralized-hybrid algorithms, respectively, as seen in Fig. \ref{percent_in_ellipse}. This indicates a realistic confidence measure for each method, in contrast to \cite{ml_rss_surya}. On the other hand, centralized-RSS and distributed-median algorithms show 97\% and 99.99\% coverage within their error ellipses. This high coverage is due to the large standard deviation of predicted coordinate estimates, reflecting a greater uncertainty in their estimates.

\subsubsection{Effect of Shadowing Noise}
\begin{figure}
\centering
\captionsetup{justification=centering}
\includegraphics[width=2.4 in]{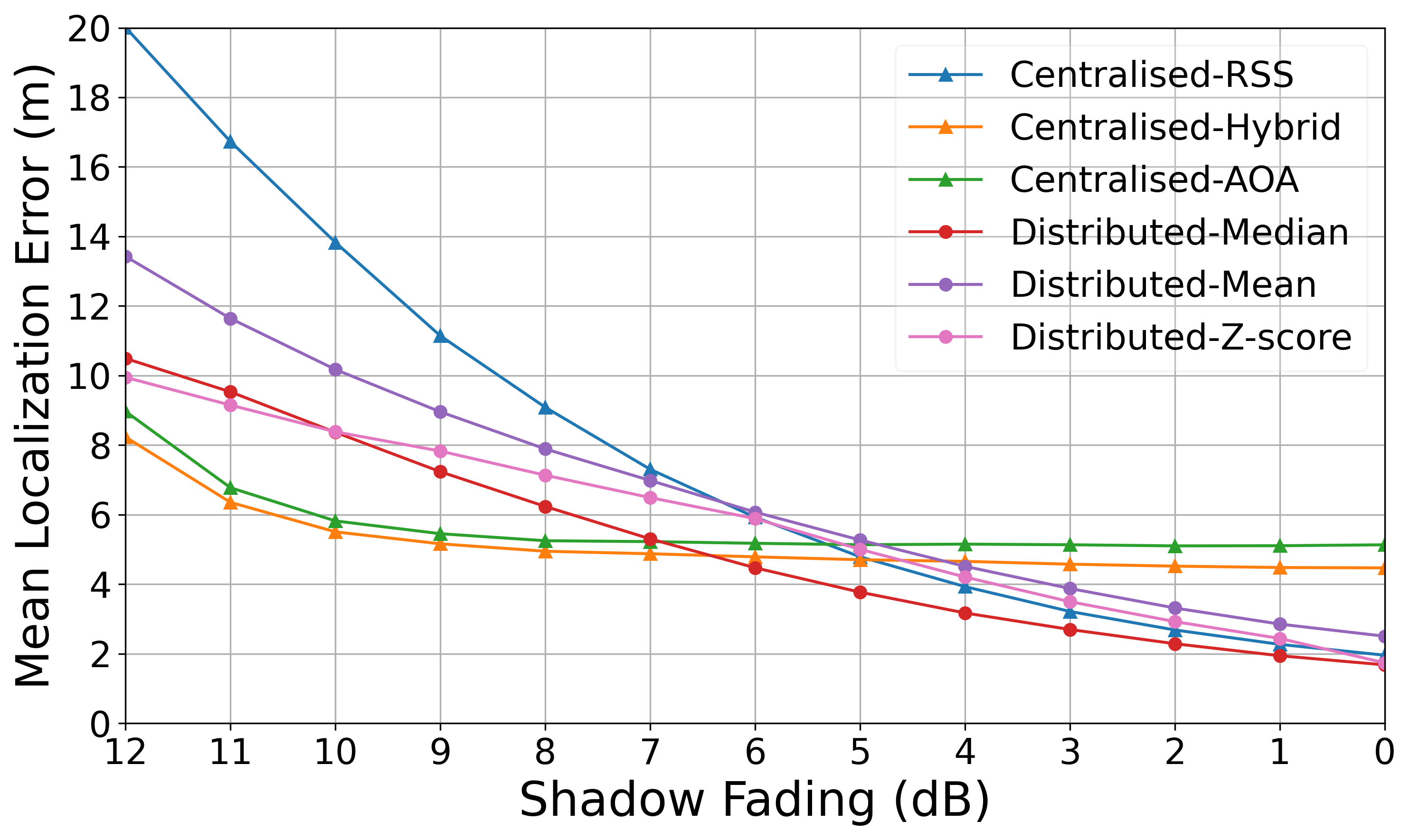}
\caption{Localization accuracy for varying shadowing noise level $\sigma_{\scriptscriptstyle SF}$ for $L=25,\ N=25$, $\mathcal{T}_z=1$, and $K=225$. }
\label{sf_vs_est_err}
\end{figure}

Fig. \ref{sf_vs_est_err} illustrates the impact of shadowing noise on the localization performance of each algorithm. For practical shadowing noise levels typically observed in outdoor cellular environments (6--10\,dB), the centralized AOA-based methods show the lowest localization errors. In contrast, the centralized-RSS method degrades significantly under high shadowing, emphasizing the sensitivity of RSS-based methods to shadowing effects. The proposed distributed algorithms perform considerably better than the conventional centralized-RSS approach. Further, we can see that among the distributed methods, median approach demonstrates the highest resilience to shadowing noise, while the z-score method achieves comparable accuracy. The 95\% error ellipse performance follows a similar trend as discussed earlier.

\subsubsection{Performance Across Varying AP Configurations}
In a cell-free massive MIMO system, increasing the number of active APs serving the UE typically leads to better SNR, thereby enhancing service quality. However, practical deployment constraints often limit the number of APs that can actively serve a UE within a given area. Therefore, localization algorithms must provide accurate results even with fewer APs, despite having less information available for ML models during the online phase.

\begin{figure}
\centering
\captionsetup{justification=centering}
\includegraphics[width=2.3 in]{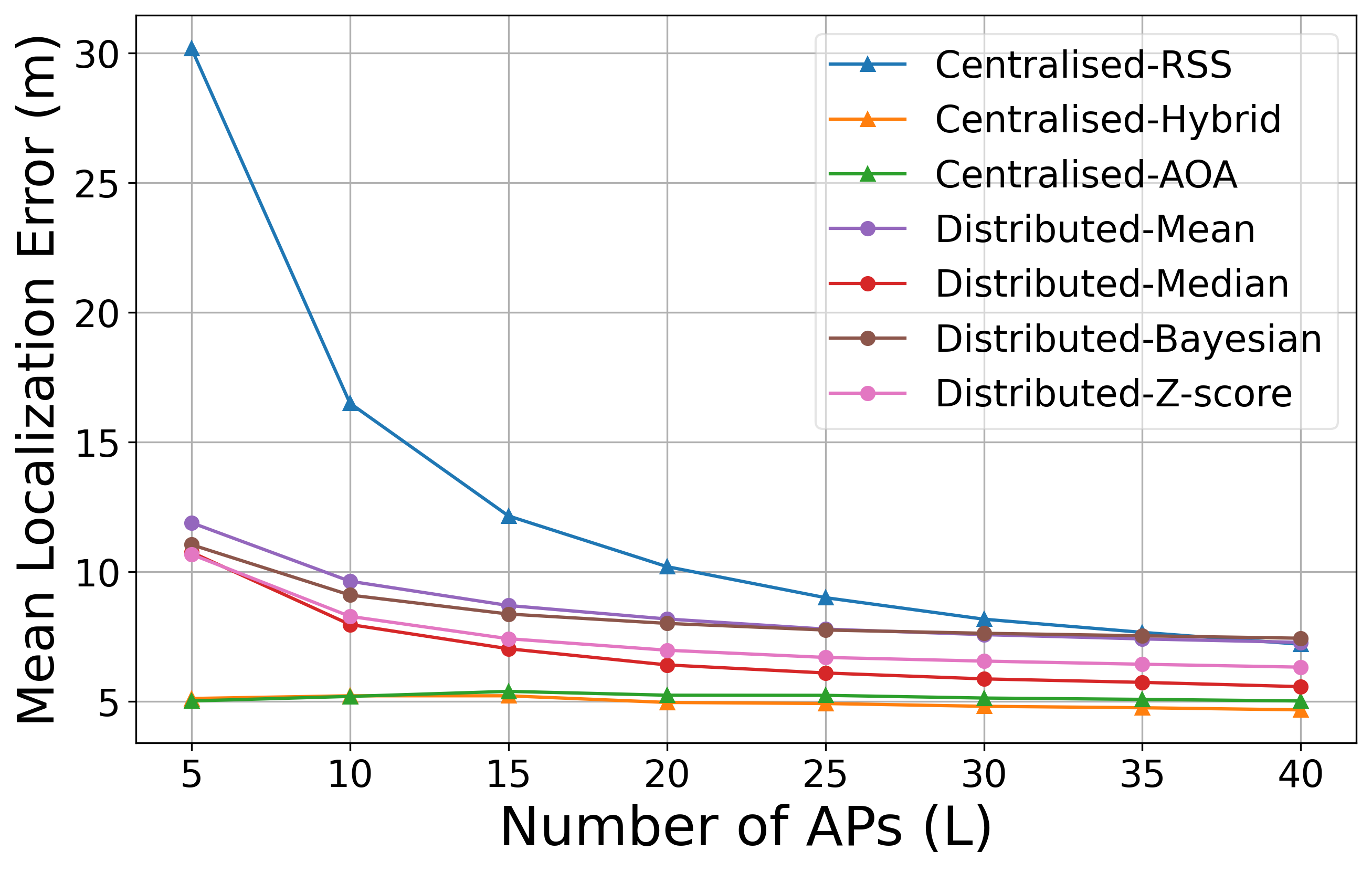}
\caption{Localization accuracy for $N$=25, $\sigma_{\scriptscriptstyle SF}$ = 8dB, $\mathcal{T}_z$=1, and $K=225$.}
\label{L_vs_est_err}
\vspace{-1.5mm}
\end{figure}

As seen in Fig. \ref{L_vs_est_err}, algorithms that use AOA inputs perform well even with as few as five APs in the simulation setup. Although centralized AOA-based algorithms generally perform best at low AP densities, distributed algorithms show comparable performance, mainly due to the reliability of AOA estimates. While the distributed approaches outperform the centralized-RSS GPR approach, they do not match the accuracy of centralized AOA and hybrid algorithms. This is because the distributed GPR models depend equally on both AOA and RSS data from each AP. If RSS data is degraded due to log-normal shadowing, the localization accuracy significantly drops. Adding more APs reduces the impact of shadowing by increasing the likelihood that some APs will experience lower shadowing, thereby improving overall performance. In contrast, centralized AOA-based algorithms aggregate AOA data from all available APs, ensuring sufficient input data for accurate localization even with fewer APs. Despite some APs having degraded RSS due to shadowing, the robust AOA information enables the centralized algorithms to provide reliable position estimates.

\subsubsection{Impact of RP Density on Localization}
\begin{figure}
\centering
\captionsetup{justification=centering}
\includegraphics[width=3.3 in]{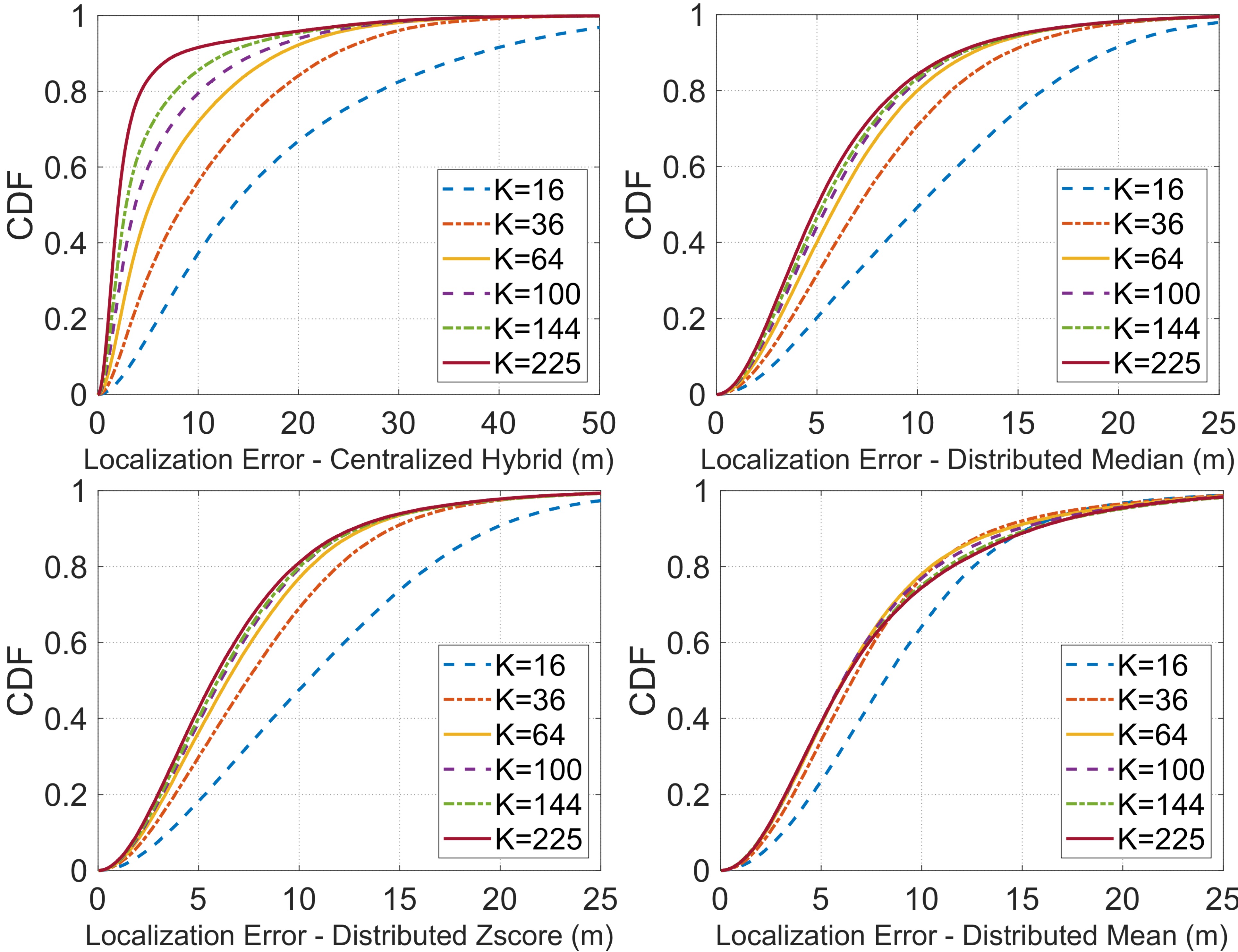}
\caption{CDF of localization error for  varying number of RPs ($K$) in the simulation area with $L=25$, $N=25$, $\sigma_{\scriptscriptstyle SF} = 8\ \text{dB}$, and $\mathcal{T}_z=1$.}
\label{5_CDF}
\vspace{-3mm}
\end{figure}

As established in \cite{my_wcl_letter}, the localization performance of centralized-hybrid GPR improves with increasing fingerprint size, since more information captured during the offline phase improves regression model training. However, the CDF for distributed algorithms, shown in Fig. \ref{5_CDF}, exhibit a slightly different trend. Specifically, for $K \geq 64$ RPs, the CDF curves converge, indicating a saturation point beyond which additional RPs offer negligible gains in localization accuracy. This suggests that the system has reached its optimal performance.

This behavior arises from the fact that each AP learns position coordinates as a function of a single RSS and AOA value. As a result, the learning process involves a relatively simple function that can be accurately modeled with fewer training points, unlike centralized algorithms that require larger training datasets. In contrast to the centralized-hybrid algorithm, the CDF saturates around a localization error of 25\,m, indicating stable test point estimates in a consistent error range without large deviations. Moreover, the saturation in localization error performance with respect to both the number of RPs and the number of antennas per AP suggests that increasing the number of APs may be the most viable strategy to further improve performance in the distributed framework.

\subsubsection{Localization Performance as AOA estimation achieves the CRB}
\begin{figure}
\centering
\captionsetup{justification=centering}
\includegraphics[width=2.5 in]{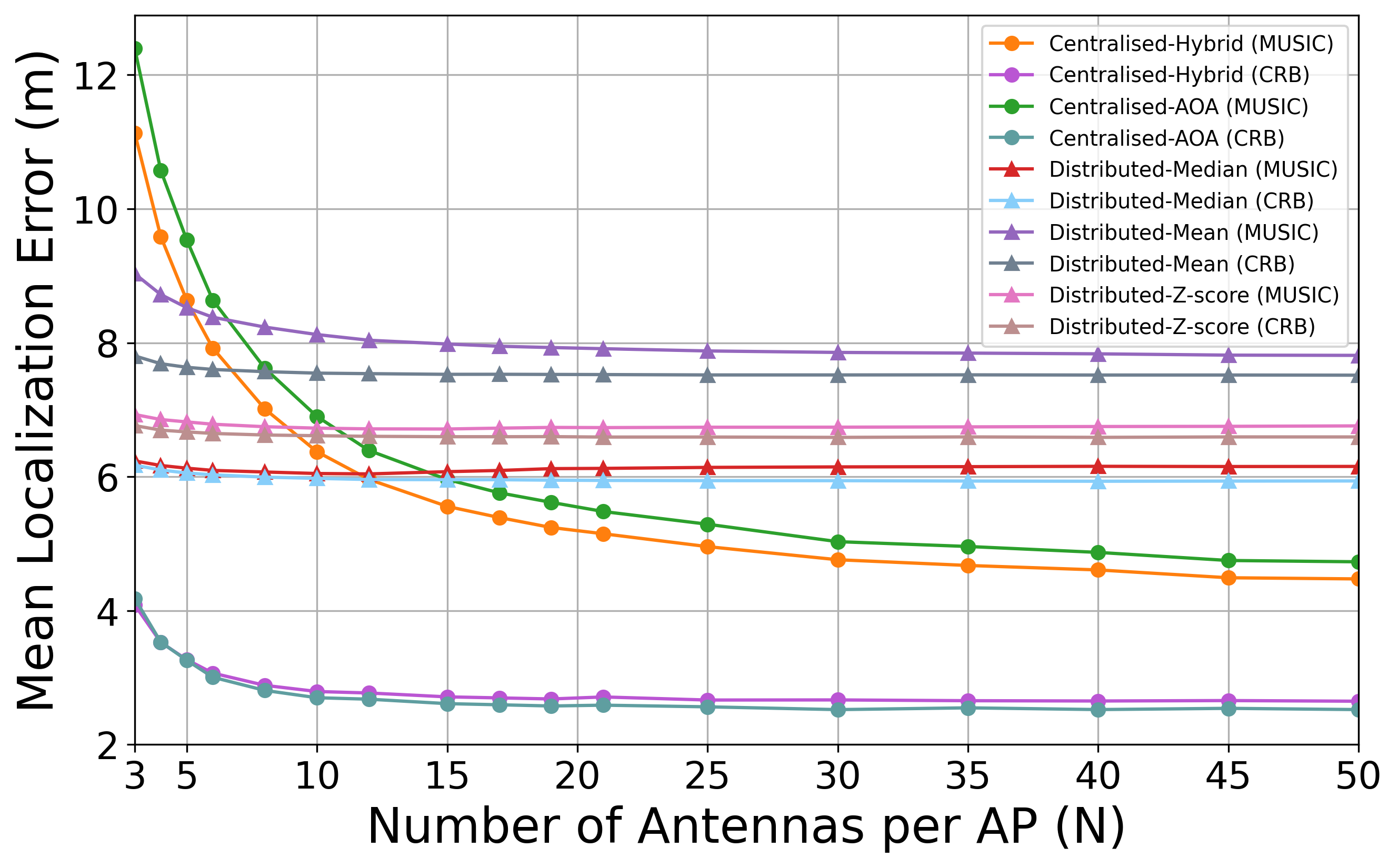}
\caption{Localization performance with CRB-achieved AOA Estimation for varying $N$, with $L=25,\ K=225,\ \sigma_{SF}=8dB$, $\mathcal{T}_z=1$.}
\label{n_vs_est_err_crb}
\vspace{-3mm}
\end{figure}

The CRB offers a fundamental limit on the variance of any unbiased estimator, providing a theoretical benchmark for the accuracy of estimation processes~\cite{class_and_mod_doa}. In the context of AOA estimation, the CRB characterizes the performance limit of direction-finding systems, by specifying the minimum achievable error variance. An error term distributed as $\mathcal{N}(0,v_{\scriptscriptstyle CRB})$ is added to the true AOA and the mean localizaton error is evaluated for various algorithms. Here, $v_{\scriptscriptstyle CRB}$ is the theoretical minimum variance quantified by the CRB (see Appendix \ref{appendixc} for derivation). When AOA estimation equals the CRB, a reduction in average localization error is observed for both centralized and distributed algorithms, compared to the use of MUSIC, as shown in Fig. \ref{n_vs_est_err_crb}. Notably, with CRB-level estimation, the localization accuracy of the centralized-AOA algorithm surpasses that of the centralized-hybrid approach, suggesting that AOA alone may provide sufficient localizaton accuracy, potentially eliminating the need for RSS inputs in centralized algorithms. This demonstrates that further improvements in localizaton quality can be achieved by employing more accurate AOA estimation methods.

\vspace{-1mm}
\section{Conclusion}
In this paper we proposed a novel low-latency localization framework for distributed cell-free massive MIMO systems, where each AP independently trains a GPR model using the local RSS and AOA data. The UE aggregates predictions from the APs to estimate the final position. This approach eliminates the need for fronthaul communication and distributes computation across APs, thereby reducing latency. Simulation results show that the proposed distributed method outperforms the centralized approach at low RP densities and achieves comparable accuracy at higher densities. Moreover, it provides improved uncertainty performance, as  indicated by the $95\%$ error ellipse. This work lays the groundwork for advancing low-latency localization in ISAC systems, encouraging future research into more efficient localization techniques for next-generation networks.
\label{conclusion}

\appendices
\vspace{\vertspacebeforesections}
\section{Statistical Distributions of Position Estimates}
\label{appendixa}
Let the $L$ normal posterior densities for the x-coordinate of the test point $x_{\scriptscriptstyle TP}$, obtained from the GPR models of $L$ APs be represented as $\mathcal{X}_{\ell} \sim \mathcal{N}(\mu_{\scriptscriptstyle 1\ell},{{v}}_{\scriptscriptstyle 1\ell})$. Since the RSS and AOA measurements from different APs are independent, due to the APs being placed far apart to ensure independent shadowing, the $L$ random variables $\mathcal{X}_{\ell}$ are also independent, as they are derived from $L$ distinct and independently trained GPR models. The sample mean of these $L$ Gaussian random variables $\mathcal{X} = \frac{1}{L}\sum\limits_{\ell=1}^{L}\mathcal{X}_{\ell}$ is also Gaussian with mean 
\begin{equation} 
\label{eqn28}
\tag{28}
\begin{split}
\mathbb{E}[{\mathcal{X}]} &= \mathbb{E}{\left[\frac{1}{L}\sum\limits_{\ell=1}^{L}\mathcal{X}_{\ell} \right]} = {\frac{1}{L}\sum\limits_{\ell=1}^{L}\mu_{\scriptscriptstyle 1\ell}}
= {\frac{1}{L}\sum\limits_{\ell=1}^{L}\overline{x}_\ell^{est}}
\end{split}
\end{equation} 
and variance
\begin{equation}
\label{eqn29}
\tag{29}
{Var}[{\mathcal{X}]} =
\resizebox{0.71\columnwidth}{!}{$
\begin{split}
 Var{\left[\frac{1}{L}\sum\limits_{\ell=1}^{L}\mathcal{X}_{\ell} \right]} 
= \frac{1}{L^2}\sum\limits_{\ell=1}^{L}Var(\mathcal{X}_{\ell}) 
= \frac{1}{L^2}\sum\limits_{\ell=1}^{L}v_{\scriptscriptstyle 1\ell}.
\end{split}
$}
\end{equation}
Thus, the components of the distributed-mean position estimate  $\overline{\boldsymbol{p}}_{\scriptscriptstyle DM}$ can be interpreted as the means of Gaussian random variables with variances $v^{\scriptscriptstyle DM}_i = \left(\frac{1}{L^2}\sum\limits_{\ell=1}^{L}v_{\scriptscriptstyle i\ell}\right)$, $i = \{1,2\}$.

Further, let $\overline{v}_i = \frac{1}{L} \sum_{\ell=1}^{L} v_{\scriptscriptstyle 1\ell}$ represent the average of $L$ GPR variance estimates. With this we obtain,
\begin{equation} 
\label{eqn30}
\tag{30}
\begin{split}
v^{\scriptscriptstyle DM}_1 = \frac{1}{L^2}\sum\limits_{\ell=1}^{L}v_{\scriptscriptstyle 1\ell} 
= \frac{\overline{v}_i}{L}.
\end{split} 
\end{equation} 
Consequently, the variance of the final coordinate estimate from the distributed-mean GPR is $L$ times smaller than the average of the estimated variances of $L$ APs.

Further, using the arithmetic mean-harmonic mean inequality, we get $v^{\scriptscriptstyle DB}_1 < v^{\scriptscriptstyle DM}_1$, implying $v^{\scriptscriptstyle DB}_1 < \frac{\displaystyle \overline{v}_i}{L}$ from (\ref{eqn30}). For the distributed-z-score GPR, 
\begin{equation} 
\label{eqn31}
\tag{31}
\resizebox{0.85\columnwidth}{!}{$
\begin{split}
v^{\scriptscriptstyle DB}_1 =  \left( \sum_{\ell=1}^L \frac{1}{v_{\scriptscriptstyle i\ell}} \right)^{-1} \leq v^{\scriptscriptstyle DZ}_1 = \left( \sum_{\ell \in \mathcal{L}_x} \frac{1}{v_{\scriptscriptstyle i\ell}} \right)^{-1} \leq \frac{\overline{v}_{\scriptscriptstyle \mathcal{L}_x}}{count_x}
\end{split} 
$}
\end{equation} 
 since $\mathcal{L}_{x} \subseteq \{1,2,\ldots,L\}$. Here $count_x$ is the cardinality of $\mathcal{L}_{x}$ and $\overline{v}_{\scriptscriptstyle \mathcal{L}_x} = \big( \sum_{\ell \in \mathcal{L}_x} v_{\scriptscriptstyle i\ell} / \text{count}_x \big)$. When only a few APs with outlier estimates are eliminated, we would have $L \approx count_x$, giving $v^{\scriptscriptstyle DB}_1 \approx v^{\scriptscriptstyle DZ}_1$. 

From a statistical perspective, the position estimate $\overline{x}_{\scriptscriptstyle DD}$ in the case of distributed-median algorithm can be interpreted as a special case of the distributed-mean algorithm, wherein only the median AP(s) contribute to the final localization estimate. Specifically, $v^{\scriptscriptstyle DD}_1 = v_{ \scriptscriptstyle 1\ell_{x}}$ for odd $L$ and $v^{\scriptscriptstyle DD}_1 = \left( 
  \frac{
    \displaystyle v_{\scriptscriptstyle 1\ell_{x\scalebox{0.4}{1}}} + 
    v_{\scriptscriptstyle 1\ell_{y\scalebox{0.4}{1}}}
  }{4}
\right)$ for even $L$. Notably, when $(L > count_x \gg 2)$, which is typically the case in cell-free massive MIMO systems, we have  $(v^{\scriptscriptstyle DB}_1 < \{v^{\scriptscriptstyle DZ}_1,v^{\scriptscriptstyle DM}_1\} \ll v^{\scriptscriptstyle DD}_1)$. Analogous reasoning applies to the variance estimates of y-coordinates.

\vspace{\vertspacebeforesections}
\section{The Error Ellipse}
\label{appendixb}
The 95\% error ellipse provides a graphical representation of the confidence region where the true location of the test point $\boldsymbol{p}_{\scriptscriptstyle TP} = (x_{\scriptscriptstyle TP} , y_{\scriptscriptstyle TP})$ is likely to be found, based on the mean of the estimates $\overline{\boldsymbol{q}}_{\scriptscriptstyle est} = (\overline{x}_{\scriptscriptstyle est},\overline{y}_{\scriptscriptstyle est})$ and their uncertainties  $\boldsymbol{v}_{\scriptscriptstyle est} = (v_1,v_2)$~\cite{err_ellipse}. For two independent normal distributions, the confidence region corresponding to a 95\% confidence level can be defined using the bivariate normal distribution as,
\begin{equation} 
\label{eqn_32}
\tag{32}
\frac{(x - \overline{x}_{\scriptscriptstyle est})^2}{v_1} + \frac{(y - \overline{y}_{\scriptscriptstyle est})^2}{v_2} \leq \chi^2_{2,0.95}.
\end{equation}
Here $\chi^2_{2,0.95}$ refers to the value of the chi-square distribution with 2 degrees of freedom that corresponds to the 95th percentile, which is approximately 5.991. The use of $\chi^2_{2}$ distribution is justified because it represents the sum of the squares of two independent standard normal variables. The equation for the 95\% error ellipse then becomes
 \begin{equation} 
\label{eqn33}
\tag{33}
\left(\frac{x - \overline{x}_{\scriptscriptstyle est}}{\sqrt{5.991 v_1}}\right)^2 + \left(\frac{y - \overline{y}_{\scriptscriptstyle est}}{\sqrt{5.991 v_2}}\right)^2 \leq 1
\end{equation}
which defines an ellipse centered at $(\overline{x}_{\scriptscriptstyle est},\overline{y}_{\scriptscriptstyle est})$, with semi-major and semi-minor axes lengths of $\sqrt{5.991 v_1}$ and $\sqrt{5.991 v_2}$ respectively, giving the area as defined in (\ref{eqn25}).

\vspace{-1mm}
\section{CRB for AOA Estimates in Disk Scattering Model}
\label{appendixc}
For $\mathzapf{S}_{\scriptscriptstyle M}$ independent measurements of the received signal $\textbf{Y}_{\scriptscriptstyle TP,\ell}$, the CRB is given by~\cite{class_and_mod_doa} $v_{\scriptscriptstyle CRB} = \mathzapf{F}^{-1}/{\mathzapf{S}_{\scriptscriptstyle M}}$
where $\mathzapf{F}$ is the Fisher information matrix, defined as
\begin{equation}
\tag{34}
\label{eqn34}
\mathzapf{F} = \text{Tr}\left( \textbf{R}_{\scriptscriptstyle TP,\ell}^{-1} \frac{\partial \textbf{R}_{\scriptscriptstyle TP,\ell}}{\partial \varphi_{\scriptscriptstyle TP,\ell}} \textbf{R}_{\scriptscriptstyle TP,\ell}^{-1} \frac{\partial \textbf{R}_{\scriptscriptstyle TP,\ell}}{\partial \varphi_{\scriptscriptstyle TP,\ell}}\right).
\end{equation}
For the disc scattering model, differentiating the expression in 
(\ref{eqn27}) yields
\begin{equation}
\tag{35}
\label{eqn35}
\resizebox{0.89\columnwidth}{!}{$
\begin{aligned} 
\frac{\partial \textbf{R}_{\scriptscriptstyle TP,\ell}}{\partial \varphi_{\scriptscriptstyle TP,\ell}} &= \rho\beta_{\scriptscriptstyle TP,\ell}\left( \textbf{G}(\zeta)\odot \dot{\textbf{a}}(\varphi_{\scriptscriptstyle TP,\ell})\textbf{a}^H(\varphi_{\scriptscriptstyle TP,\ell}) \right. \\
& \left. +\ \dot{\textbf{G}}(\zeta)\odot\textbf{a}(\varphi_{\scriptscriptstyle TP,\ell})\textbf{a}^H(\varphi_{\scriptscriptstyle TP,\ell}) + \textbf{G}(\zeta)\odot \textbf{a}(\varphi_{\scriptscriptstyle TP,\ell})\dot{\textbf{a}}^H(\varphi_{\scriptscriptstyle TP,\ell}) \right),
\end{aligned}
$}
\end{equation}
where
\begin{equation}
\tag{36}
\label{eqn36}
\resizebox{0.89\columnwidth}{!}{$
\begin{split}
[\dot{\textbf{G}}(\zeta)]_{mn} &= \frac{\partial [{\textbf{G}}(\zeta)]_{mn}}{\partial \varphi_{\scriptscriptstyle TP,\ell}} = \frac{\partial J_0\left((m-n)\zeta\right)}{\partial \varphi_{\scriptscriptstyle TP,\ell}}  
+ \frac{\partial J_2\left((m-n)\zeta\right)}{\partial \varphi_{\scriptscriptstyle TP,\ell}} \\
&= \frac{2\pi d}{\lambda}(m-n)\Delta sin(\varphi_{\scriptscriptstyle TP,\ell})J_1((m-n)\zeta) \\
& \qquad - \frac{2\pi d}{\lambda}(m-n)\Delta \frac{sin(\varphi_{\scriptscriptstyle TP,\ell})}{2}\Bigl( J_1((m-n)\zeta) - J_3((m-n)\zeta)\Bigr) \\
&= \frac{\pi d}{\lambda}(m-n)\Delta sin(\varphi_{\scriptscriptstyle TP,\ell})\Bigl(J_1((m-n)\zeta) + J_3((m-n)\zeta)\Bigr)
\end{split} 
$}
\end{equation}
and $\dot{\textbf{a}}(\varphi_{\scriptscriptstyle TP,\ell}) = \frac{\partial \textbf{a}(\varphi_{\scriptscriptstyle TP,\ell})}{\partial \varphi_{\scriptscriptstyle TP,\ell}} = -j\frac{2\pi d}{\lambda}cos(\varphi_{\scriptscriptstyle TP,\ell})\textbf{D}\textbf{a}(\varphi_{\scriptscriptstyle TP,\ell})$, $\textbf{D} = \text{diag}\{[0,1,2,\dots,N-1]\}$.

\nocite{}
\bibliographystyle{IEEEtranDOI}
\bibliography{IEEEabrv,bibLatex}
\end{document}